\documentclass[twocolumn,showpacs,amsmath,amssymb]{revtex4}
\input{epsf}

\usepackage{graphicx}
\usepackage{array}
\usepackage{longtable}
\usepackage{rotating,booktabs}
\usepackage{booktabs,threeparttable}
\usepackage{bm}
\usepackage{hyperref}

\usepackage{color}

\begin{document}

\title{Magic wavelengths of Ca$^{+}$ ion for linearly and circularly polarized light}

\author{Jun Jiang}
\email {phyjiang@yeah.net}
\author{Li Jiang}
\author{Xia Wang}
\author{Deng-Hong Zhang}
\author{Lu-You Xie}
\author{Chen-Zhong Dong}

\affiliation{Key Laboratory of Atomic and Molecular 
Physics and Functional Materials of Gansu Province,
College of Physics and Electronic Engineering, 
Northwest Normal University, Lanzhou 730070, P. R. China
}

\date{\today}

\begin{abstract}
The dynamic dipole polarizabilities of 
the low-lying states of Ca$^{+}$ for linearly and circularly polarized light 
are calculated by using relativistic configuration interaction plus core 
polarization (RCICP) approach.  
The magic wavelengths, at which the two levels of the transitions have the 
same ac Stark shifts, for $4s$-$4p_{j,m}$ 
and $4s$-$3d_{j,m}$ magnetic sublevels transitions are determined. The present magic wavelengths 
for linearly polarized light agree with the available results excellently. 
The polarizability for the circularly polarized light has 
the scalar, vector and tensor components.
The dynamic polarizability is different for
each of magnetic sublevels of the atomic state. 
Additional magic wavelengths have been found for the circularly polarized light.  
We recommend that the measurement of the magic wavelength near 850 nm for 
$4s-4p_{\frac32,m=\pm\frac32,\pm\frac12}$ could be able to determine the oscillator strength ratio of
$f_{4p_{\frac32} \to 3d_{\frac32}}$ and $f_{4p_{\frac32} \to 3d_{\frac52}}$.
\end{abstract}

\pacs{31.15.ac, 31.15.ap, 34.20.Cf} \maketitle

\section{INTRODUCTION}

The magic wavelength, at which the ac Stark shift of the transition energy is zero for the certain
frequencies, was introduced in Refs.\cite{ye99a, katori99b}. 
The magic wavelengths have been extensively used in ultraprecise 
optical lattice clocks\cite{takamoto03a,bauch03a,gill03a,gill05a,lorini08a,gill11a,kirchmair09a},
the state-insensitive quantum engineering \cite{sahoo13a,wilpers07a}.

The magic wavelengths for the linearly polarized light have been studied  
for alkali-metal and alkaline-earth-metal atoms in experiment 
and theory\cite{ludlow08a,kaur15a,liu15a,tang13b,lundblad10a,arora07a,singh16b}.
The magic wavelengths of alkali-metal atoms 
for the circularly polarized light have 
been calculated\cite{arora12a,sahoo13a,singh16a,cs16a}. 
The use of circularly polarized light has advantages 
owing to the vector polarizabilities which are absent in the linearly polarized light, 
such as magnetic-sublevel selective 
trapping and far-off-resonance laser trapping\cite{kien13a,sahoo13a}. 

Ca$^+$ is an alkali-metal like ion. 
Since the nuclear spin is zero for $^{40}$Ca$^{+}$, it is immune to 
the first-order Zeeman frequency shift\cite{kelin13a} and convenient for laser cooling. 
The $^{40}$Ca$^{+}$ ion is preferred for optical frequency standard and 
quantum computing\cite{hertz12a,degenhardt04a,chwalla09a,zhang16a,haff08a}. 
The frequency of $^{40}$Ca$^{+}$ optical clocks has been measured
with an uncertainty at the $10^{-17}$ level \cite{huang16a}.

Recently, two magic wavelengths of the $^{40}$Ca$^{+}$ $4s-3d_{\frac52,m=\frac12,\frac32}$ clock 
transitions for the linearly polarized light
are measured with very high precision \cite{liu15a}. This measurement agrees with  
the theoretical predictions of B-spline Dirac-Fock plus core 
polarization (DFCP) method \cite{tang13b} excellently. Meanwhile,  
the measurement for these two magic wavelengths determines the ratio of 
$4s-4p_{\frac12}$ and $4s-4p_{\frac32}$  oscillator strength 
with the deviation less than 0.5\%. 

In this paper, the energy levels, electric dipole matrix elements 
and static polarizabilities are calculated using 
relativistic configuration interaction plus core polarization (RCICP) method.
The dynamic dipole polarizabilities of the
$4s$, $4p_{j}$ and $3d_{j}$ states of Ca$^{+}$ ion are calculated for
the linearly and circularly polarized light.
The magic wavelengths for each of magnetic 
sublevel transitions are determined. 
In Sec. II., a brief description of the theoretical method
is presented. In Sec. III., the static and dynamic polarizabilities 
and magic wavelengths are discussed.  
In Sec. IV., a few conclusions are pointed out.
The unit used in the present calculations is atomic
unit (a.u.).
\begin{table*}                                                                                                                          
\caption{\label{tab1}Relativistic dipole polarizabilities (a.u.) of the ground and low-lying excited states of Ca$^{+}$ ion. 
The numbers in parentheses are uncertainties by introducing 0.5\% uncertainties
into the dominant matrix elements. }         
\begin{ruledtabular}                                                                                       
\begin{tabular}{lllllllll}  
&\multicolumn{1}{c}{$4s_{\frac12}$}& \multicolumn{2}{c}{$3d_{\frac32}$}& \multicolumn{2}{c}{$3d_{\frac52}$} &\multicolumn{1}{c}{$4p_{\frac12}$} & \multicolumn{2}{c}{$4p_{\frac32}$}	\\
\cline{3-4} \cline{5-6} \cline{8-9}
& \multicolumn{1}{c}{$ \alpha_{1}$} &\multicolumn{1}{c}{$ \alpha_{1}$} & \multicolumn{1}{c}{$ \alpha^{t}_{1}$}
&\multicolumn{1}{c}{$ \alpha_{1}$} & \multicolumn{1}{c}{$ \alpha^{t}_{1}$} & \multicolumn{1}{c}{$ \alpha_{1}$}
&\multicolumn{1}{c}{$ \alpha_{1}$} & \multicolumn{1}{c}{$ \alpha^{t}_{1}$} \\
\hline 
\multicolumn{9}{c}{Dipole} \\                                                    
Present                     &  75.46(72)  &   32.98(24) &  $-$17.97(17) &    32.80(24) &  $-$25.28(24) &   $-$2.98(11)  &    $-$1.12(10)  &  10.20(11)    \\    
DFCP \cite{tang13b}         &  75.28      &   32.99     &  $-$17.88     &    32.81     &  $-$25.16     &   $-$2.774     &   $-$ 0.931     &  10.12     \\    
MBPT-SD \cite{safronova11a} & 76.1(5)     &   32.0      &  $-$17.43(23) &    32.0      &  $-$24.51(29) &   $-$0.75      &       1.02(64)  &  10.31(28) \\    
CICP \cite{mitroy08a}       &  75.49      &   32.73     &  $-$17.64     &    32.73     &  $-$25.20     &   $-$2.032     &   $-$2.032      &  10.47     \\    
RCC  \cite{sahoo09a}        &  73.0(1.5)  &   28.5(1.5) &  $-$15.87     &    29.5(1.0) &  $-$22.49(5)  &                &                 &  \\    
RCC-STO \cite{sahoo09a}     &  74.3       &   31.6      &  $-$17.7      &    32.5      &  $-$25.5      &                &                 & \\    
f-sums\cite{chang83a}       &  75.3(4)    &             &               &              &               &                &                 &   \\   
\end{tabular}                                                                                                                          
\end{ruledtabular}                                                                                                                     
\end{table*}   
\begin{table*}                                                                                                                          
\caption{\label{tab2}Comparison of static scalar polarizabilities (a.u.)
for $4s_{\frac12}$ and $3d_{\frac52}$ states and blackbody radiation shift (Hz) for the 
$4s_{\frac12}- 3d_{\frac52}$ transition of $^{40}$Ca$^{+}$ ion at T=300 K.} 
\begin{ruledtabular}                                                                                       
\begin{tabular}{lllllllll} 
   & Present  & Ref.\cite{mitroy08a} & Ref.\cite{sahoo09a} & Ref.\cite{arora07a} & Ref.\cite{cham04a} & Ref.\cite{kajita05a} & Expt.\cite{gao16a} & Ref.\cite{huang12a,yaohuang11a} \\
\hline
$\alpha_{0}(4s_{\frac12})$    & 75.46(72)   & 75.49         & 73.0(1.5)    & 76.1(1.1)   & 76        & 73   & 76.1(1.1)   &  \\
$\alpha_{0}(3d_{\frac52})$    & 32.80(24)   & 32.73         & 29.5(1.0)    & 32.0(1.1)   & 31        & 23   & 31.8(3)     & \\
$\Delta \alpha$               & 42.66       & 42.76         & 43.5         & 44.1        & 45        & 40   & 44.3        &   \\
\multicolumn{1}{c}{$\eta(4s_{\frac12})$ }   & $-3.0[-9]$    &              &             &           &      &             & & \\
\multicolumn{1}{c}{$\eta(3d_{\frac52})$}    & $-1.1[-8]$    &              &             &           &      &             & & \\
BBR shift                     & 0.367(39)   & 0.368         &  0.37(1)     & 0.38(1)     & 0.39(27)  & 0.4  & 0.35(0.009)\footnotemark[1]    & 0.35 \\
\end{tabular}                                                                                                                          
\end{ruledtabular}    
\footnotetext[1]{The temperature is 294.4 K in this experiment. }
\end{table*}
\begin{table*}                                                                  
\caption{\label{tab3} Magic wavelengths (in nm) for the transitions of 
low-lying states of Ca$^{+}$ for the linearly polarized light.
The numbers in the parentheses are uncertainties calculated by assuming certain matrix elements have
$\pm2$\% uncertainties.}          
\begin{ruledtabular}
\begin{tabular}{llllll}
 Resonance & $\lambda_{res}$ & Present & Ref.\cite{kaur15a}  & Ref.\cite{tang13b} & Exp.\cite{liu15a} \\
\cline{2-6}	
\multicolumn{1}{c}{Transitions}  & \multicolumn{4}{c}{$4s_{\frac12}-4p_{\frac12}$}           \\ 
\hline  
 $4p_{\frac12}-3d_{\frac32}$ & 866.214 & & & &                                                \\
                             &         & 691.24(12.29)   &  697.65  & 690.817(11.984)  &       \\ 
 $4p_{\frac12}-4s_{\frac12}$ & 396.847 & & & &                                                 \\
                             &         & 395.1788(377)   &  395.18  & 395.1807(14)     &        \\   
 $4p_{\frac12}-5s_{\frac12}$ & 370.603 & & & &                                                  \\
                             &         & 368.0221(1412)  &  368.10  & 368.0149(901)    &         \\  
 $4p_{\frac12}-4d_{\frac32}$ & 315.887 & & & &                                                   \\
\multicolumn{6}{c}{$4s_{\frac12}-4p_{\frac32},m=\frac32 $}    \\  
\hline
 $4p_{\frac32}-3d_{\frac52}$ & 854.209 & & & &                 \\     
                             &         & 850.9217(15)     & 850.12   &    &   \\   
 $4p_{\frac32}-3d_{\frac32}$ & 849.802 & & & &                 \\                    
                             &         & 672.89(15.33)    & 678.35   &  672.508(11.3150) &    \\  
 $4p_{\frac12}-4s_{\frac12}$ & 396.847 & & & &                           \\                     
                             &         & 395.7729(19)     & 395.77   &  395.774(10)      &    \\  
 $4p_{\frac32}-4s_{\frac12}$ & 393.366 & & & &                 \\                      
\multicolumn{6}{c}{$4s_{\frac12}-4p_{\frac32},m=\frac12 $}    \\   
\hline
 $4p_{\frac32}-3d_{\frac52}$ & 854.209 & & & &                 \\                          
                             &         & 850.1164(1)    & 850.12   &    &       \\ 
 $4p_{\frac32}-3d_{\frac32}$ & 849.802 & & & & \\	                          
                             &         & 687.51(10.33)  & 693.76   & 687.022(12.285)   &     \\ 
$4p_{\frac12}-4s_{\frac12}$ & 396.847 & & & &  \\	                        
                            &         & 396.2297(218)  & 396.23   & 396.2315(13)      &    \\
$4p_{\frac32}-4s_{\frac12}$ & 393.366 & & & &  \\ 
$4p_{\frac32}-5s_{\frac12}$ & 373.690          \\              
                            &         & 369.6523(1849) & 369.72   & 369.6393(1534)    &    \\  
$4p_{\frac32}-4d_{\frac32}$ & 318.128 & & & & \\                        
\multicolumn{6}{c}{$4s_{\frac12}-3d_{\frac32},m=\frac32$}     \\    
\hline                      
                             &         & 887.28(3.52)  & 884.54   & 887.382(3.196)   &   \\   
 $3d_{\frac32}-4p_{\frac12}$ & 866.214 & & & & \\ 
 $4s_{\frac12}-4p_{\frac12}$ & 396.847 & & & & \\                          
                             &         & 395.7951 (1)  & 395.79   & 395.7970(1)      &     \\    
$4s_{\frac12}-4p_{\frac32}$  & 393.366 & & & & \\                       
 \multicolumn{6}{c}{$4s_{\frac12}-3d_{\frac32},m=\frac12$}    \\  
\hline                        
                             &         & 1307.60(96.2)  & 1252.44  & 1308.590(71.108) &    \\  
 $3d_{\frac32}-4p_{\frac12}$ & 866.214 & & & & \\         
                             &         & 850.3301(18)   & 850.33   & 850.335(2)       &      \\ 
 $3d_{\frac32}-4p_{\frac32}$ & 849.802 & & & & \\	
 $4s_{\frac12}-4p_{\frac12}$ & 396.847 & & & & \\                       
                             &         & 395.7962(1)    & 395.80   & 395.7981(1)      &      \\ 	
 $4s_{\frac12}-4p_{\frac32}$  & 393.366 & & & & \\                        
\multicolumn{6}{c}{$4s_{\frac12}-3d_{\frac52},m=\frac52$}     \\ 
\hline 
 $4s_{\frac12}-4p_{\frac12}$ & 396.847 & & & & \\ 
                             &         & 395.7949(1)   & 395.79    & 395.7968(1)      &   \\
  $4s_{\frac12}-4p_{\frac32}$  & 393.366 & & & & \\ 	                        
\multicolumn{6}{c}{$4s_{\frac12}-3d_{\frac52},m=\frac32$}      \\ 
\hline                          
                              &         & 1073.80(31.61)  & 1052.26  & 1074.336(26.352)  &  \\ 	
 $3d_{\frac52}-4p_{\frac32}$ & 854.209 & & & & \\
 $4s_{\frac12}-4p_{\frac12}$ & 396.847 & & & & \\                        
                              &         & 395.7958(1)  & 395.79   & 395.7978(1)  & 395.7992(2)     \\
 $4s_{\frac12}-4p_{\frac32}$  & 393.366 & & & & \\                         
\multicolumn{6}{c}{$4s_{\frac12}-3d_{\frac52},m=\frac12$}      \\     
\hline                      
                              &         & 1337.30(115.38)  & 1271.92  & 1338.474(82.593)  &              \\ 
$3d_{\frac52}-4p_{\frac32}$ & 854.209 & & & & \\
 $4s_{\frac12}-4p_{\frac12}$ & 396.847 & & & & \\ 	                        
                             &         & 395.7963(1)  & 395.79   & 395.7982(1)  & 395.7990(2)     \\ 	
$4s_{\frac12}-4p_{\frac32}$  & 393.366 & & & & \\                         	  
\end{tabular}                       
\end{ruledtabular}               
\end{table*}   

\section{Theoretical method}

The energy levels and transition arrays are calculated 
using RCICP method which has been developed recently \cite{jiang16a}.
The present method is similar to the calculation of magic wavelengths 
of Ca$^+$ for the linearly polarized light by Tang \emph{et al.} \cite{tang13b}, 
except that they use the B-spline basis.
The basic strategy of the model is to partition the atom 
into valence and core electrons.
The first step involves a Dirac-Fock (DF) 
calculation of the Ca$^{2+}$ ground state.
The orbitals of the core are written as linear combinations
of S-spinors which can be treated as relativistic 
generalizations of the Slater-type orbitals.

Then, the effective interaction of the valence electron with the
core is written as
\begin{eqnarray}        
{\bf \emph{H}} = c \bm{\alpha} \cdot \bm{p}+ \beta c^2+V_{core}(\bm{r}).
\end{eqnarray}
 The core operator is
\begin {eqnarray}
  V_{core}(\bm{r})=-\frac{Z}{r}+V_{dir}(\bm{r})+V_{exc}(\bm {r})+V_{p}(\bm{r}),
\end{eqnarray}
where $ V_{dir}$, $V_{exp}$ are direct and exchange 
interactions with core electrons. 
The semi-empirical core polarization potential is introduced to approximate 
the correlation interaction between the core and valence
electrons. The $\ell$-dependent polarization potential $V_{p}$ can be written as
\begin{eqnarray}
	V_{p}(r)=\sum\limits_{\ell m}\frac{\alpha g_{\ell}^{2}(r)}{2r^{4}}|\ell m\rangle \langle{\ell m|},
\end{eqnarray}
$\alpha$ is the static dipole polarizability of the core, $\alpha$ = 3.26 a.u.\cite{safronvoa11a}
and $g_{\ell}^{2}(r)$ is a cutoff function,
$g_{\ell}^{2}(r)=1-exp(-r^{6}/\rho_{\ell}^{6})$. $\rho_{\ell}$ is an adjustable parameter that is tuned to reproduce the 
binding energies of the corresponding states. 
The orbitals of the valence electron are written as linear combinations
of L-spinors and S-spinors. L-spinors can be treated as relativistic 
generalizations of the  Laguerre -type orbitals.
See Supplemental Tables I and II for lists of  energy levels and electric-dipole matrix elements for 
some low-lying excited states transitions of Ca$^{+}$ ions \cite{suppmat}.

For arbitrary polarized light, the dynamic polarizability 
$\alpha_{i}(\omega)$ is given by \cite{kien13a,nl86a,beloy09a}
\begin{eqnarray}
\label{a}
\alpha_{i}(\omega) &=& \alpha^{0}_{i}(\omega) + 
Acos\theta_{k}\frac{m_{j_i}}{2j_i} \alpha^{1}_{i}(\omega) \nonumber \\
&+& (\frac{3cos^{2}\theta_{p}-1}{2})\frac{3m^{2}_{j_i}-j_i(j_i+1)}{j_i(2j_i-1)} \alpha^{2}_{i}(\omega),
\end{eqnarray}
where $\alpha_i^{0}(\omega)$, $\alpha_i^{1}(\omega)$, $\alpha_i^{2}(\omega)$ are the scalar, 
vector, and tensor polarizabilities for state $i$, respectively;
$j$ ,$m_{j_i}$ are the total angular momentum and the corresponding magnetic quantum number.
$\theta_{k}$ is the angle between the wave vector of the electric field and $z$-axis.
$\theta_{p}$ relates to the direction of polarization vector and 
quantization axis which was defined in Ref.\cite{kien13a,nl86a}.
From geometrical considerations, it is found that $\theta_{k}$ and $\theta_{p}$
must satisfy the inequality $cos^2\theta_{k} + cos^2\theta_{p} \leq 1$.
A represents the degree of polarization, which can be taken arbitrary value from $-1$ to $1$. 
For the linearly polarized light, A is equal to zero. The direction of 
$z$-axis is polarization direction which is 
perpendicular to the wave vector of the electric field, $cos\theta_{k}$ = 0 and 
$cos\theta_{p}$ = 1. 
\begin{eqnarray}
\alpha_{i}(\omega) = \alpha^{0}_{i}(\omega) + 
\frac{3m^{2}_{j_i}-j_i(j_i+1)}{j_i(2j_i-1)} \alpha^{2}_{i}(\omega).
\end{eqnarray}
$A = 1$ is for the right handed and $A = -1$ is 
for the left handed circularly polarized light.
Here, we can choose the direction of 
$z$-axis as the wave vector of the electric field and then $cos\theta_{k}$ = 1, $cos^2\theta_{p}$ = $0$.
Then the eq.(\ref{a}) can be simplified as 
\begin{eqnarray}
\alpha_{i}(\omega) = \alpha^{0}_{i}(\omega) + 
A\frac{m_{j_i}}{2j_i} \alpha^{1}_{i}(\omega)  -\frac{3m^{2}_{j_i}-j_i(j_i+1)}{2j_i(2j_i-1)} \alpha^{2}_{i}(\omega).
\end{eqnarray}
The scalar polarizability was conventionally expressed as
\begin{eqnarray}
\alpha^{0}_{i}(\omega) &=& \frac{1}{3(2j_{i}+1)}\sum_{in}|\langle\psi_{i}\|D\|\psi_{n}\rangle|^{2} \nonumber \\
&\times&\left[ \frac{1}{\Delta E_{in}+\omega}+\frac{1}{\Delta E_{in}-\omega}\right]. 
\end{eqnarray}
The vector polarizability was written as
\begin{eqnarray}
\alpha^{1}_{i}(\omega) &=& -\sqrt {\frac{6j_{i}}{(j_i+1)(2j_i+1)}}\sum_{in} 
\left \{
\begin{array}{ccc}
j_i & 1    & j_n \\
1   & j_i  & 1 \\
\end {array}
\right \}\nonumber \\
&\times& (-1)^{j_i +j_n +1}|\langle\psi_{i}\|D\|\psi_{n}\rangle|^{2} \nonumber \\ 
&\times&\left[ \frac{1}{\Delta E_{in}+\omega}-\frac{1}{\Delta E_{in}-\omega}\right].
\end{eqnarray}
The tensor polarizability was expressed as
\begin{eqnarray}
\alpha^{2}_{i}(\omega) &=& -2\sqrt {\frac{5j_{i}(2j_{i}-1)}{(6(j_{i}+1)(2j_{i}+1)(2j_{i}+3)}} \nonumber \\
& \times & \sum_{in} \left \{
\begin{array}{ccc}
j_i & 1& j_n \\
1 & j_i  & 2 \\
\end {array}
\right \}  (-1)^{j_{i} +j_{n} +1}|\langle\psi_{i}\|D\|\psi_{n}\rangle|^{2} \nonumber \\ 
&\times&\left[ \frac{1}{\Delta E_{in}+\omega}+\frac{1}{\Delta E_{in}-\omega}\right].
\end{eqnarray}
where $|\langle\psi_{i}\|D\|\psi_{n}\rangle|$ and $\Delta E_{in}$ are 
reduced matrix element and excitation energy of transition respectively.
For the state with $j<\frac12$, the tensor polarizability makes no contribution to total polarizability.
\begin{figure*}[tbh]
\label{fig1}
\centering{
\includegraphics[width=17cm]{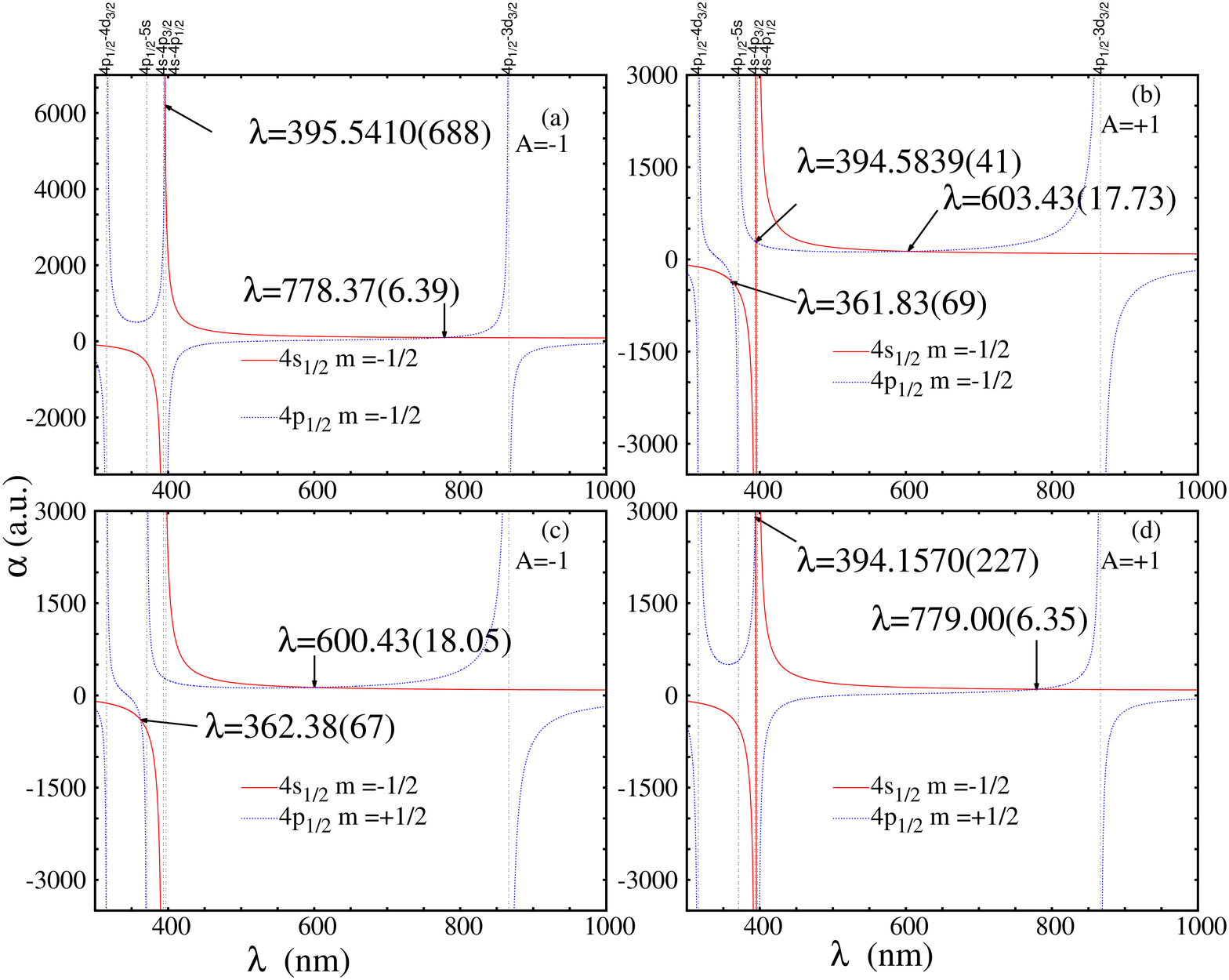}
} \caption{ \label{fig1} (color online) Dynamic polarizabilities (in au) for the   
$4s_{\frac12m=-\frac12}$ and $4p_{\frac12m=\pm\frac12}$ states of Ca$^{+}$ for the left and right 
handed circularly polarized light. The various magic wavelengths are identified by arrows.
The vertical lines identify the resonance transition wavelengths.}
\end{figure*}
\begin{figure*}[tbh]
\centering{
\includegraphics[width=17cm]{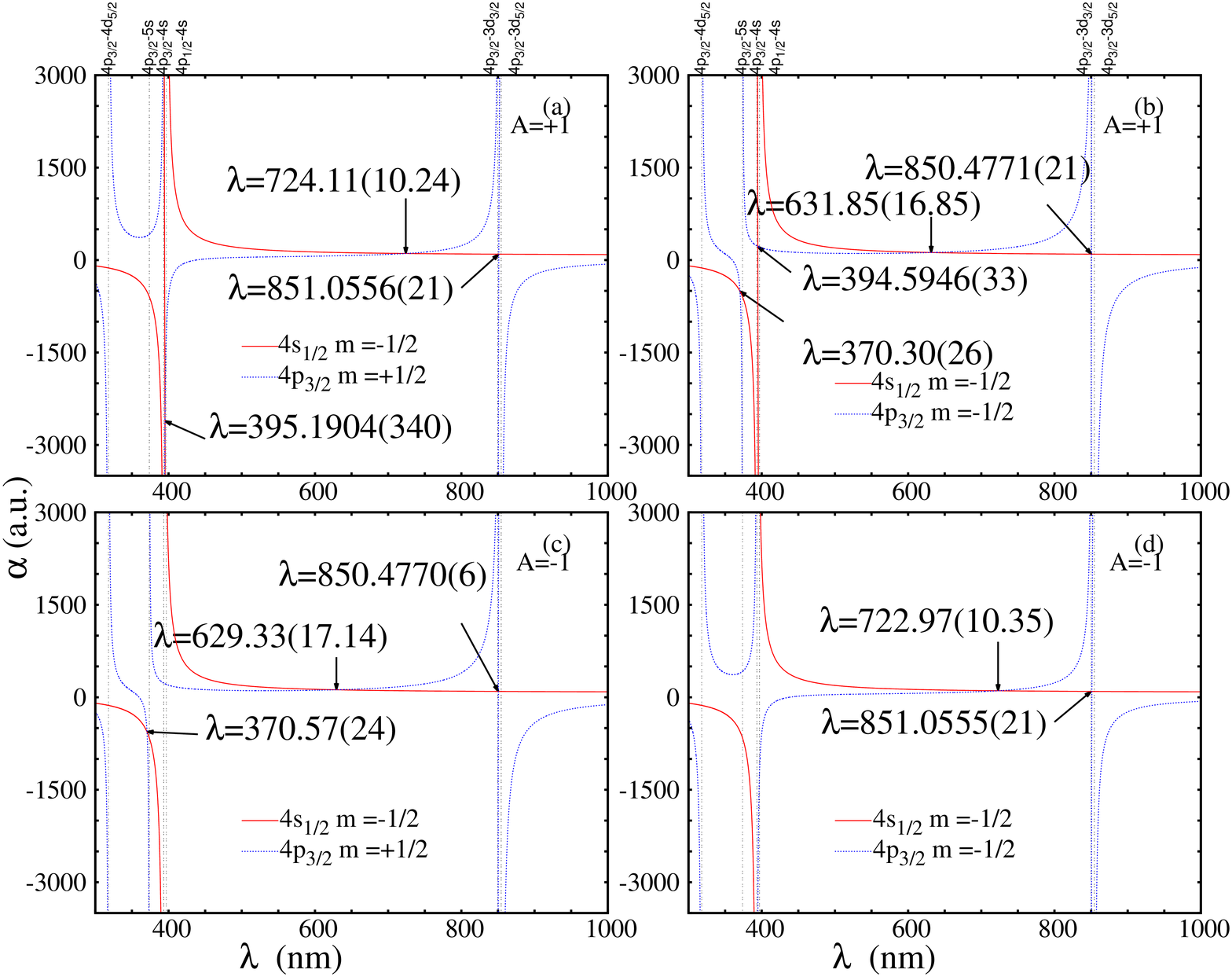}
} \caption{ \label{fig2} (color online) Dynamic polarizabilities (in au) 
for the $4s_{\frac12m=-\frac12}$ and $4p_{\frac32m=\pm\frac12}$ states
 of Ca$^{+}$ for the left and right handed circularly polarized 
light.The various magic wavelengths are identified by arrows. 
The vertical lines identify the resonance transition wavelengths.}
\end{figure*}
\begin{figure*}[tbh]
\centering{
\includegraphics[width=17cm]{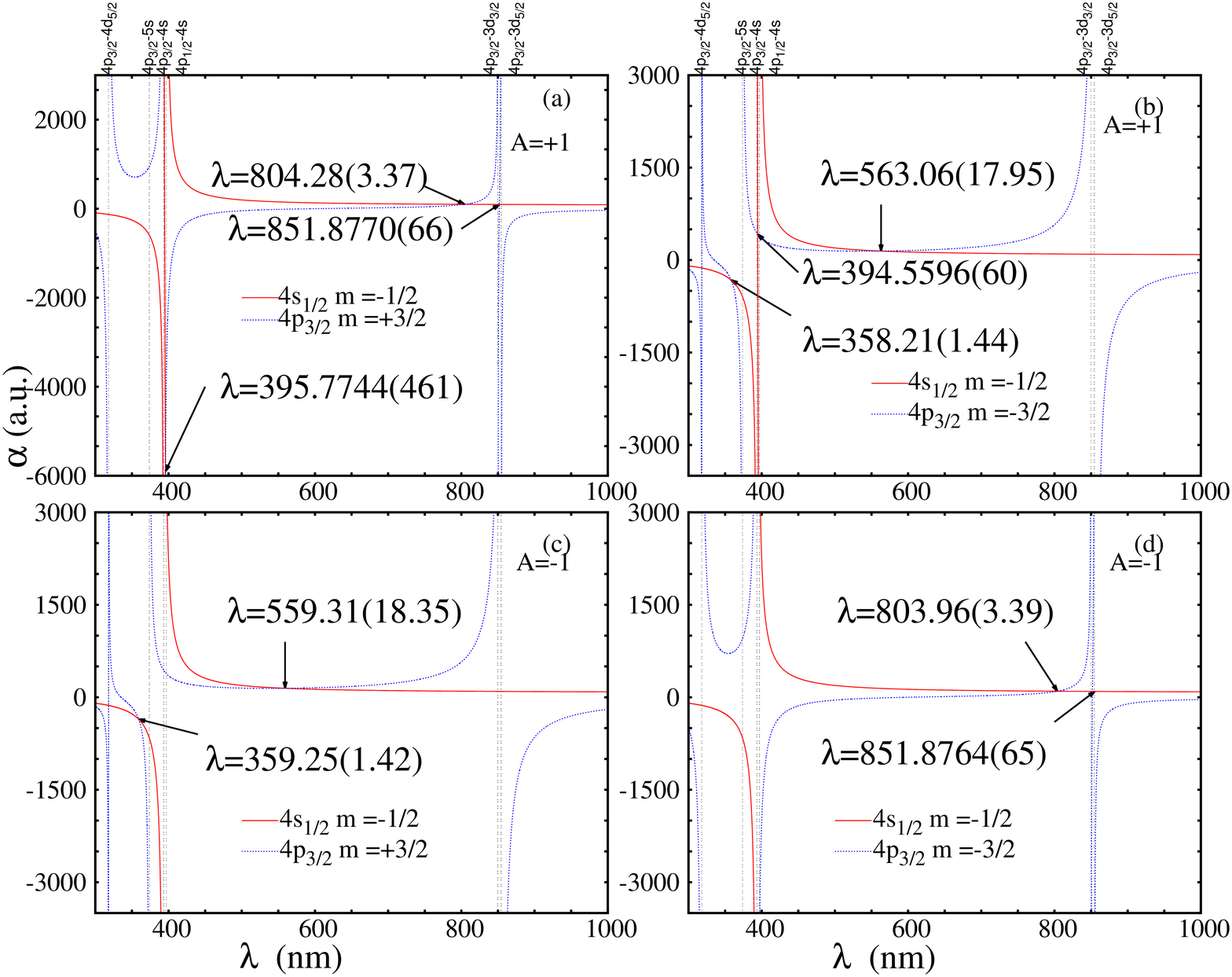}
} \caption{ \label{fig3} (color online) Dynamic polarizabilities 
(in au) for the  $4s_{\frac12m=-\frac12}$ and $4p_{\frac32m=\pm\frac32}$ states
 of Ca$^{+}$ for the left and right handed circularly polarized light. 
The various magic wavelengths are identified by arrows. 
The vertical lines identify the resonance transition wavelengths.}
\end{figure*}
\begin{figure}[tbh]
\centering{
\includegraphics[width=9.cm]{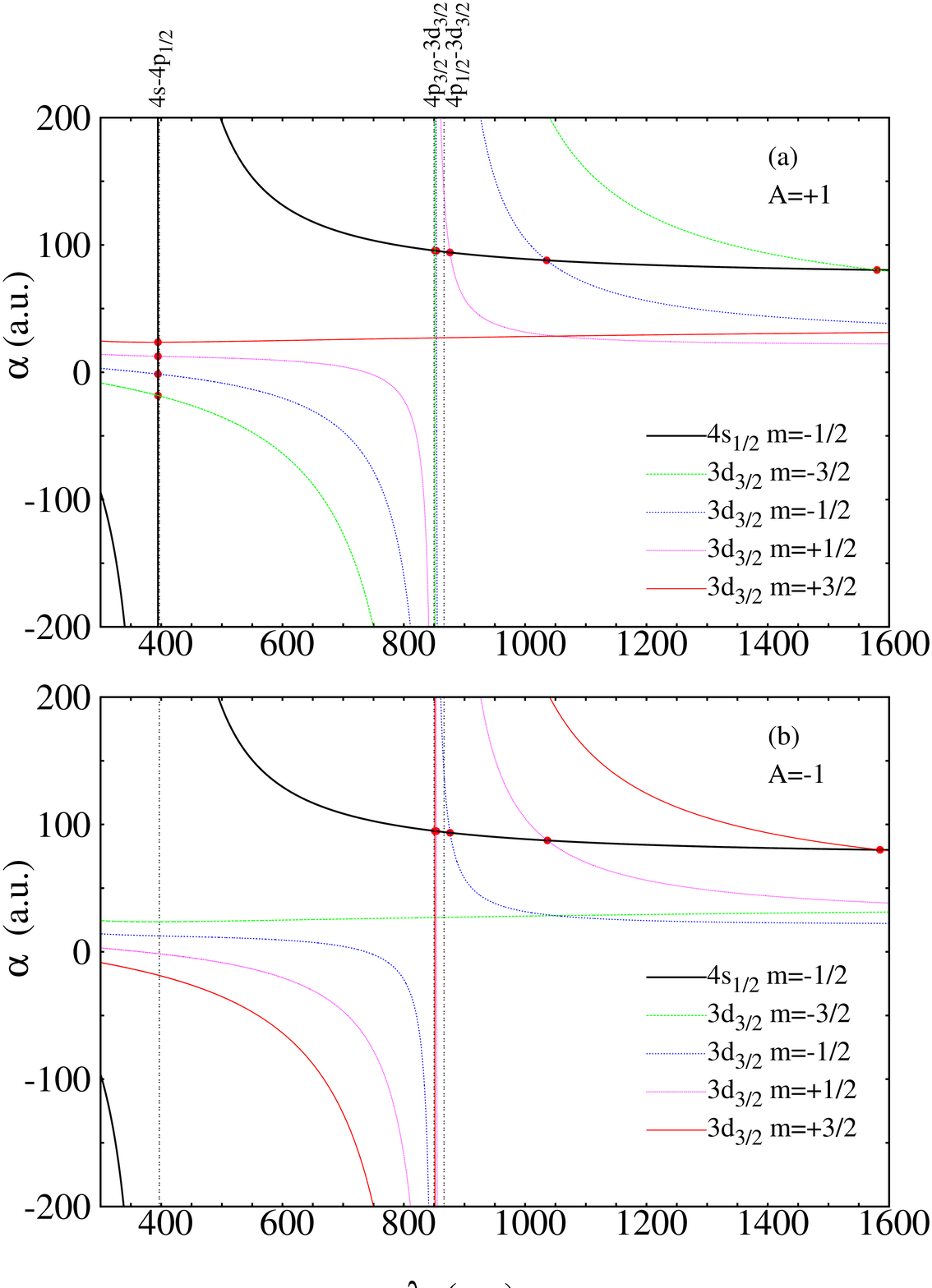}
} \caption{ \label{fig4} (color online) Dynamic polarizabilities (in au) 
for the $4s_{\frac12m=-\frac12}$ and $3d_{\frac32m_j}$ states
 of Ca$^{+}$ for the left and right handed circularly polarized light. 
The various magic wavelengths are identified by red points. 
The vertical lines identify the resonance transition wavelengths. }
\end{figure}
\begin{figure}[tbh]
\centering{
\includegraphics[width=9cm]{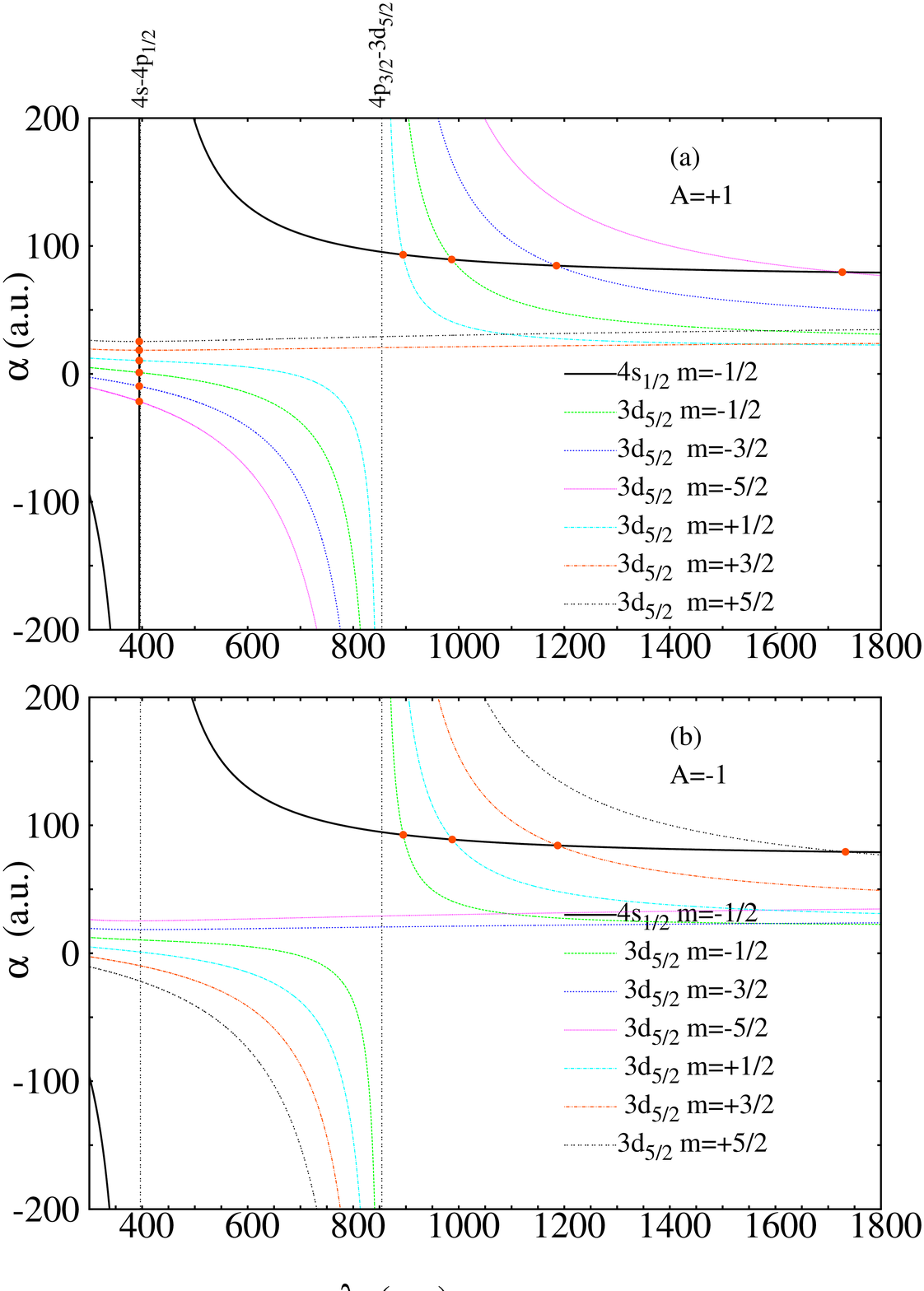}
} \caption{ \label{fig5} (color online) Dynamic polarizabilities (in au) 
for the $4s_{\frac12m=-\frac12}$ and $3d_{\frac52m_j}$ states
 of Ca$^{+}$ for the left and right handed circularly polarized light. 
The various magic wavelengths are identified by red points.
The vertical lines identify the resonance transition wavelengths. }
\end{figure}

\section{ POLARIZABILITIES }

\subsection{STATIC POLARIZABILITIES }

When the frequency of light is zero, the dynamic polarizability
becomes static dipole polarizability, in which the contribution of vector polarizability is zero.
Table \ref{tab1} lists dipole scalar and tensor static polarizabilities of the 
low-lying states of Ca$^{+}$, which are compared with the 
available theoretical and experimental results. 

For the ground state, the present polarizabilities are in good agreement with 
the previous calculations\cite{tang13b,mitroy08a,chang83a,safronova11a}.
The errors are within 0.8\% except for the value calculated by relativistic coupled cluster (RCC) method.
There is a significant difference between the values of RCC and other calculations.
The scalar and tensor polarizabilities of the $3d_j$ states are excellently 
agree with the results of Tang \emph{et al.} \cite{tang13b}. 
The scalar polarizabilities for the $4p_j$ states are negative which are caused by the 
downward transitions to the lower states.

\subsection{BLACKBODY RADIATION SHIFT}

The accuracy of optical frequency standards is limited by the frequency shift in the 
clock transitions caused by the interactions of the ion with external field.
The major contributions to the systematic frequency shifts come from 
blackbody radiation (BBR) shift.
The frequency shift for a state due to 
blackbody radiation at temperature T can be written as\
\begin{eqnarray}
\Delta\nu=-\frac{1}{2}(831.9V/m)^{2}(\frac{T(K)}{300})^{4}\alpha_i^{0}(1+\eta),
\end{eqnarray}
where $\alpha_i^{0}$ is the static scalar polarizability for state $i$. 
The factor $\eta$ is a small dynamic correction \cite{porsev06a},
\begin{eqnarray}
\eta &=&\frac{(80/63)\pi^{2}}{\alpha^{(E1)}_{n}(0)T}\sum_{i}
\frac{|\langle\psi_{i}\|D\|\psi_{n}\rangle|^{2}}{(2J_{n}+1)y_{i}^{3}}  \nonumber \\  
&\times&(1+\frac{21\pi^{2}}{5y_{i}^{2}}+\frac{336\pi^{4}}{11y_{i}^{4}}).
\end{eqnarray}
As Table \ref{tab2} shows the value of $\eta$ can be negligible.
Here $y_{i}=\omega_{in}/T$ and T is temperature, we set as 300 K. 
The atomic unit for $\alpha$ can be converted to SI units via 
$\alpha/h [Hz/(V/m)^{2}]=2.48832\times 10^{-8}\alpha [a.u.]$.

The BBR shift for the clock transition is 
the difference of the BBR shifts between 
the individual levels involved in the transition and 
can be written as,
\begin{eqnarray}
\Delta\nu_{BBR}(4s-3d_{\frac52})=\Delta\nu(3d_\frac52)-\Delta\nu(4s).
\end{eqnarray}
Table \ref{tab2} shows the BBR shift and scalar dipole polarizabilities of 
individual levels for $4s_{\frac12}-3d_{\frac52}$ clock transition of Ca$^{+}$ ion.
The present BBR shift is $0.367$ Hz which agrees with the value of 
Mitroy \emph{et al.}\cite{mitroy08a} about 0.16\%,
Sahoo \emph{et al.}\cite{sahoo09a} about 0.7\%,
Arora \emph{et al.}\cite{arora07a} about 3.3\%,
Champenois \emph{et al.}\cite{cham04a} about 5.7\%
and Kajita \emph{et al.}\cite{kajita05a} about 8.15\%.
In the recent experiment, the BBR shift is 0.35(0.009) Hz at  
temperature T= $294.4\pm1.6$ K\cite{gao16a}. 
If we set T = 294.4 K in our calculation, the BBR shift is 0.341 Hz which is within 
the experimental error bar. 

\subsection{  DYNAMIC POLARIZABILITIES }

\subsubsection{linearly polarized light}

For the case of the linear polarization (A=0), the vector polarizability 
does not contribute to total polarizability.
The polarizability only has scalar component for state with $j=\frac12$. 
The polarizability has scalar and tensor components for state with $j>\frac12$.
Supplemental Fig.I gives dynamic polarizabilties of $4s$, $4p_j$ and $3d_j$ states  
for the linearly polarized light\cite{suppmat}.

Table \ref{tab3} lists magic wavelengths for the 
$4s_{\frac12}-4p_{\frac12,\frac32}$ and $4s_{\frac12}-3d_{\frac32,\frac52}$ transitions of Ca$^{+}$ 
for the linearly polarized light. 
Uncertainties for all the magic wavelengths have been estimated which are similar 
to the estimations of Tang \emph{et al.}\cite{tang13b}.
For the $4s-4p_j$ polarizability difference, the matrix elements of $4s-4p_j$,
$4p_j-5s$, $4p_j-3d_j$ and $4p_j-4d_j$ are dominant.
For the $4s-3d_j$ polarizability difference, the matrix elements of $4s-4p_j$ and
$3d_j-4p_j$ are dominant. All these matrix elements were changed by $\pm2\%$ and the magic wavelengths
were recomputed.
The magic wavelengths near 368 nm, 395 nm and 850 nm for $4s-4p_j$ and $4s-3d_j$ transitions
agree with the results
of available theoretical values\cite{tang13b,kaur15a} excellently. 
The maximum difference is 0.7 nm at 850.9217 nm for $4s-4p_{\frac32}$ transition.
Compared to experimental value \cite{liu15a},
the differences of magic wavelengths at 395.7958 nm and 395.7963 nm  for $4s-3d_{\frac52}$, 
are only 0.0034 nm and 0.0027 nm, respectively.
Present magic wavelengths near 691 nm, 672 nm, 
687 nm, 887 nm, 1073 nm and 1307 nm for $4s-4p_j$ and 
$4s-3d_j$ transitions are in 
good agreement with results of Tang \emph{et al.}\cite{tang13b}.
The maximum difference is 0.99 nm near 1307 nm for $4s-3d_{\frac32m=\frac12}$ transition.  
But the results of Kaur \emph{et al.}\cite{kaur15a} have big 
differences with the present results and 
Tang's results \cite{tang13b} for these magic wavelengths, 
For example, the differences are 6.4 nm near 691 nm 
for the $4s_{\frac12} - 4p_{\frac12}$ transition, 55 nm near 1307.60 nm
for $4s_{\frac12} - 3d_{\frac32}$ transition. 
Supplemental Tables III-X lists the breakdown of the polarizabilities at the magic 
wavelengths\cite{suppmat}.
               
\subsubsection{circularly polarized light}

For the circularly polarized light, the polarizability has 
the scalar, vector and tensor components.
The dynamic polarizability is different for
each of magnetic sublevels of the atomic state. 
Due to the symmetry, the dynamic polarizabilities of 
negative $m$ state for $A=-1$ will be same as positive $m$ state for $A=+1$.
In the following discussion, we just give the polarizabilites for the ground state $4s_{\frac12m=-\frac12}$ .

Fig.\ref{fig1} shows the dynamic polarizabilities 
of the $4s_{\frac12 m=-\frac12}$ and $4p_{\frac12 m=\pm\frac12}$ states 
for the left and right handed circularly polarized light. 
In Fig. \ref{fig1}(a), we find that the dynamic polarizabilities 
of $4s_{\frac12 m=-\frac12}$ state have a big difference
with the dynamic polarizabilities 
for the linearly polarized light.
When the wavelength is close to $4s-4p_{\frac12}$ transition wavelength, 
$\alpha_{4s}$ $\rightarrow$ $\infty$ for linearly polarized light.
However, the dynamic polarizabilities  of $4s_{\frac12m=-\frac12}$ state are not infinity for 
left handed circularly polarized light.
The reason is that the $4s-4p_{\frac12}$ transition has no 
contribution to the polarizability 
when the photo energy is close to the $4s-4p_{\frac12}$ transition energy.
The scalar and vector polarizabilities offset each other, that is
\begin{eqnarray}
\label{z}
 & & \lim_{\omega \rightarrow \Delta E(4s-np_{\frac12})} 
( \frac{1}{3(2j_{i}+1)}|\langle\psi_{4s}\|D\|\psi_{np_{\frac12}}\rangle|^{2}  \nonumber \\
&\times&\left[ \frac{1}{\Delta E_{4s-np_{\frac12}}+\omega}+\frac{1}{\Delta E_{4s-np_{\frac12}}-\omega}\right] \nonumber \\
& - & A\frac{m_{j_i}}{2j_i} \sqrt {\frac{6j_{i}}{(j_i+1)(2j_i+1)}} 
\left \{
\begin{array}{ccc}
j_i & 1    & j_n \\
1   & j_i  & 1 \\
\end {array}
\right \}\nonumber \\
&\times& (-1)^{j_i +j_n +1}|\langle\psi_{4s}\|D\|\psi_{np_{\frac12}}\rangle|^{2} \nonumber \\ 
&\times&\left[ \frac{1}{\Delta E_{4s-np_{\frac12}}+\omega}-\frac{1}{\Delta E_{4s-np_{\frac12}}-\omega}\right] ) \nonumber \\
&=&  0.
\end{eqnarray}
where, $j_i$ = $\frac12$ is total angular momentum for 4s state,
$j_n$ is total angular momentum for $np_{\frac12}$ state
and $m_{j_i}$ = $-\frac12$ is magnetic quantum number for $4s$ state.
Same situations also happened for $4p_{\frac12m =-\frac12}$ state when  
the wavelength is close to $4p_{\frac12}-5s$ transition wavelength.
There are two magic wavelengths for $4s_{\frac12m=-\frac12}-4p_{\frac12m=-\frac12}$ transition for A=$-1$. 
The first magic wavelength is 395.5410 nm which lies between $4s-4p_{\frac12}$
and $4s-4p_{\frac32}$ resonances energy. This magic wavelength is
very close to the magic wavelength 395.1788 nm for $4s-4p_{\frac12}$ transition 
for linearly polarized light \cite{tang13b, kaur15a}. 
Another magic wavelength 778.37 nm occurs 
between $4p_{\frac12}-4s$ and $4p_{\frac12}-3d_{\frac32}$ transition energy.
This magic wavelength has 87 nm difference with the magic wavelength
691.24 nm of $4s-4p_{\frac12}$ transition for linearly polarized light which also lies  
between $4p_{\frac12}-4s$ and $4p_{\frac12}-3d_{\frac32}$ transitions.

Fig.\ref{fig1}(b) gives the dynamic polarizabilities of $4s_{\frac12m=-\frac12}$
and $4p_{\frac12m=-\frac12}$ for right handed circularly polarized light.
The change of polarizabilities of $4s$ state is similar to linearly polarized light. 
Three magic wavelengths are found for $4s_{\frac12m=-\frac12}-4p_{\frac12m=-\frac12}$ transition.
The first magic wavelength 361.83 nm occurs  between the energies of
the $4p_{\frac12}-4d_{\frac32}$ and the $4p_{\frac12}-5s$ transitions.
The second magic wavelength 394.5839 nm occurs between $4s-4p_{\frac12}$
and $4s-4p_{\frac32}$ resonant transition. 
The next wavelength is located around 603 nm when the photo energy
gets to close to the excitation energies between the $4p_{\frac12}-4s$ and
$4p_{\frac12}-3d_{\frac32}$ transitions.

Fig.\ref{fig1}(c) gives the dynamic polarizabilities of $4s_{\frac12m=-\frac12}$
and $4p_{\frac12m=+\frac12}$ for left handed circularly polarized light.
Only two magic wavelengths 362.38 and 600.43 nm are found in the considered range of wavelength.
The first one occurs between $4p_{\frac12}-4d_{\frac32}$ and $4p_{\frac12}-5s$ transitions.
Another one  occurs when the photo energy
lies between $4p_{\frac12}-3d_{\frac32}$ and $4p_{\frac12}-4s$ transition energies.
There is no magic wavelength between $4s-4p_{\frac12}$ and $4s-4p_{\frac32}$ resonant transition.

Fig.\ref{fig1}(d) gives the dynamic polarizabilities of $4s_{\frac12m=-\frac12}$
and $4p_{\frac12 m=+\frac12}$ for right handed circularly polarized light.
Two magic wavelengths are found for $4s_{\frac12m=-\frac12}-4p_{\frac12 m=\frac12}$ transition.
The first magic wavelength is 394.1570 nm which lies between $4s-4p_{\frac12}$
and $4s-4p_{\frac32}$ resonant transition.
The second magic wavelength is located 779.00 nm when the photo energy lies between 
the $4p_{\frac12}-3d_{\frac32}$ and $4p_{\frac12}-4s$ transitions. 

Fig.\ref{fig2} shows the dynamic polarizabilities 
of the $4s_{\frac12m=-\frac12}$ and $4p_{\frac32m=\pm\frac12}$ states for 
the left and right handed circularly polarized light. 
Fig.\ref{fig2}(a) shows the dynamic polarizabilities of 
$4s_{\frac12m=-\frac12}$ and $4p_{\frac32m=\frac12}$ states for A=$+1$.
There are three magic wavelengths in the range of 360 nm to 1000 nm for this transition.
There is no magic wavelength between $4p_{\frac32}-4d_{\frac52}$ and $4p_{\frac32}-5s$ transitions.
The first magic wavelength 395.1904 nm lies between the $4s-4p_{\frac12}$ 
and $4s-4p_{\frac32}$ resonant transition, that just has 1.0 nm 
difference compared to that for the linearly polarized light.
The second magic wavelength 724.11 nm lies between 
the $4p_{\frac12}-4s$ and $4p_{\frac32}-3d_{\frac32}$ transitions, which 
is larger about 37 nm than the 687.51 nm for linearly polarized light.
The last one 851.0556 nm is close to the magic wavelength 
850.1164 nm for linearly polarized light which lies between the transition energy 
of $4p_{\frac32}-3d_{j}$.  

Much more attentions should be paid on the magic wavelength near 851 nm, 
because this wavelength arises due to 
cancellations in the polarizabilties from two transitions of
$4p_{\frac32} \to 3d_{j}$ spin-orbital splitting. From the Table XXV in Suppmlement\cite{suppmat}, it can 
be found that the  $4s_{\frac12m=-\frac12}$ polarizability is dominated by $4s_{\frac12} - 4p_j$
transitions and the $4p_{\frac32m=\frac12}$ polarizabiltiy 
is dominated by $4p_{\frac32} \to 3d_{j}$ transitions.
Combining with the experimental matrix elements of $4s \to 4p_j$
transitions, the measurement of this magic wavelength
could be able to determine the oscillator strength ratio of
$f_{4p_{\frac32} \to 3d_{\frac32}} : f_{4p_{\frac32} \to 3d_{\frac52}}$. Suppose that all
the remaining components accuracy of $4p_{\frac32}$ polarizability
including the $4p_{\frac32} \to 5s_{\frac12}$ and $4d_j$ contributions is $5\%$. Then the overall
uncertainty to the polarizability is less than $1\%$. 

The dynamic polarizabilities of $4s_{\frac12m=-\frac12}$  and $4p_{\frac32m=-\frac12}$ states for A=$+1$
are shown in Fig.\ref{fig2}(b).
There are four magic wavelengths in the range of 360 nm to 1000 nm.
The first magic wavelength 370.30 nm occurs when the photo energy lies between energy of 
$4p_{\frac32}-4d_{\frac52}$ and $4p_{\frac32}-5s$ transitions. This magic wavelength is 
larger about 0.6 nm than the magic wavelength 
369.65 nm for $4s-4p_{\frac32m=\frac12}$ transition for linearly polarized light,
which also lies between $4p_{\frac32}-4d_{\frac52}$ and $4p_{\frac32}-5s$ transitions.
The second magic wavelength is 395.5946 nm which lies between the
$4s-4p_{\frac12}$ and $4s-4p_{\frac32}$ resonant transition.
The third one is 631.85 nm when photo energy lies between
the $4p_{\frac32}-3d_{\frac52}$ and $4p_{\frac12}-4s$ transitions.
The last magic wavelength occurs 850.4771 nm which occurs between
the $4p_{\frac32}-3d_{\frac32}$ and $4p_{\frac32}-3d_{\frac52}$ resonant transition.
\begin{table}                                                                                                                         
\caption{\label{tab4} Magic wavelengths (in nm) 
for the $4s_{\frac12m=-\frac12 }-3d_{\frac32m_j}$  transition 
of Ca$^{+}$ with the circularly polarized lights.} 
\begin{ruledtabular}                                                                                                                  
\begin{tabular}{llll} 
\multicolumn{1}{c}{A=-1} & \multicolumn{1}{c}{A=1}&\multicolumn{1}{c}{A=-1} & \multicolumn{1}{c}{A=1}	\\  	
\cline{1-2}\cline{3-4}                                                                        
\multicolumn{2}{c}{$m_{j}=\frac32$}& \multicolumn{2}{c}{$m_{j}=-\frac32$}\\                      
1584.95(143.70)   &		           &          & 1580.01(142.74)\\
851.1728(42)      &	                   &          & 851.1724(42)   \\                                                                    
	          &	 394.6315(2)       &          & 394.6394(4)    \\ 
\multicolumn{2}{c}{$m_{j}=\frac12$}        &  \multicolumn{2}{c}{$m_{j}=-\frac12$} \\                    
1036.70(20.30)    &                        &                &   1035.46(20.15) \\                                                                      
853.5998(264)     & 876.07(2.59)           & 876.28(2.61)   &   853.5974(266) \\                                                                      
		  & 394.6335(1)            &                &   394.6362 (1)\\                                                                       
\end{tabular}                                                                                                                         
\end{ruledtabular}                                                                                                                    
\end{table}                                                                                                                           
\begin{table}                                                                                                                         
\caption{\label{tab5} Magic wavelengths (in nm) for the $4s_{\frac12m=-\frac12}-3d_{\frac52mj}$ 
transition of Ca$^{+}$ with the circularly polarized lights.}                                           
\begin{ruledtabular}                                                                                                                  
\begin{tabular}{cccc}                                                                                                                
\multicolumn{1}{c}{A=-1} & \multicolumn{1}{c}{A=1}&\multicolumn{1}{c}{A=-1} & \multicolumn{1}{c}{A=1}	\\  	
\cline{1-2}\cline{3-4}   	   			
\multicolumn{2}{c}{$m_{j}=\frac52$}& \multicolumn{2}{c}{$m_{j}=-\frac52$}	\\                    	                                                                  1732.77(200.04) &                &            & 1726.68(198.7)\\                                                                     
		&   394.6311(3)  &            &  394.6400(4)  \\                                                                    
\multicolumn{2}{c}{$m_{j}=\frac32$}& \multicolumn{2}{c}{$m_{j}=-\frac32$}  \\                                                                                               1187.42(46.79) &            &            &  1185.07(46.46)    \\                                                                 
&  394.6324(2)  &             &  394.6377(2)        \\                                                              
\multicolumn{2}{c}{$m_{j}=\frac12$}& \multicolumn{2}{c}{$m_{j}=-\frac12$} \\                                                                                                987.60(14.87)   &  894.50(4.03) &  894.81(4.06)   &  986.63(14.76)    \\                                                                   
	        & 394.6339(1)  &                 &  394.6357(1)    \\                                                                   
\end{tabular}                                                                                                                         
\end{ruledtabular}                                                                                                                    
\end{table}
The dynamic polarizabilities of $4s_{\frac12m=-\frac12}$  
and $4p_{\frac32m=+\frac12}$ states for $A=-1$
are shown in Fig.\ref{fig2}(c). 
Three magic wavelengths are got for 
the $4s_{\frac12m=-\frac12}-4p_{\frac32m=+\frac12}$ transition.
The first one 370.57 nm lies  between the 
$4p_{\frac32}-4d_{\frac52}$ and $4p_{\frac32}-5s$ transitions.
The second one 629.33 nm lies between 
the $4p_{\frac32}-3d_{\frac52}$ and $4p_{\frac12}-4s$ transitions,
that has about 58 nm difference compared to 687.51 nm for the linearly polarized light.
The last one is 850.4770 nm which 
lies between $4p_{\frac32}-4d_{j}$ transitions.

The dynamic polarizabilities of $4s_{\frac12m=-\frac12}$  
and $4p_{\frac32m=-\frac12}$ states for A=$-1$
are shown in Fig.\ref{fig2}(d).
The only two magic wavelengths are found.
The first one 722.97nm lies between the 
$4p_{\frac32}-3d_{\frac52}$ and $4p_{\frac32}-4s$ transitions.
Another one 851.0555 nm occurs between the
$4p_{\frac32}-3d_{\frac52}$ and $4p_{\frac32}-3d_{\frac32}$ transitions.

Fig.\ref{fig3} shows the dynamic polarizabilities 
of the $4s_{\frac12m=-\frac12}$  and $4p_{\frac32m=\pm\frac32}$ states for the left and right handed 
circularly polarized light. 
The dynamic polarizabilities of $4s_{\frac12m=-\frac12}$  and $4p_{\frac32m=+\frac32}$ states for A=+1
are shown in Fig.\ref{fig3}(a).
There are three magic wavelengths in the range of 370 nm to 1000 nm.
The first magic wavelength 395.7744 nm occurs when the photo energy lies between energy of 
$4s-4p_{j}$ resonant transitions. 
The second magic wavelength 804.28 nm which lies between 
$4p_{\frac32}-4s$ and $4p_{\frac32}-3d_{\frac32}$  transitions 
is larger about 131 nm than 672.89 nm for the linearly polarized light.
The third magic wavelength 851.8770 nm is close to the magic wavelength 850.9217 nm
for the linearly polarized light.
Three magic wavelengths are found for the 
$4s_{\frac12m=-\frac12} - 4p_{\frac32m=-\frac32}$ transition for the right handed polarized light
in Fig.\ref{fig3}(b), 
that are 358.21 nm 394.5596 nm and 563.06 nm.
The dynamic polarizabilities of $4s_{\frac12m=-\frac12}$  and $4p_{\frac32m=+\frac32}$ states for A=-1
are shown in Fig.\ref{fig3}(c). 
Two magic wavelengths for this transition
are found at 359.25 nm  and 559.31 nm.
The dynamic polarizabilities of $4s_{\frac12m=-\frac12}$  and $4p_{\frac32m=-\frac32}$ states for A=-1
are shown in Fig.\ref{fig3}(d).
Two magic wavelengths 803.96 nm and 851.8764 nm are found. 
                        
The dynamic polarizabilities of the $4s_{\frac12m=-\frac12}$  and $3d_{\frac32m}$ states are shown in Fig.\ref{fig4}.
We also find that the dynamic polarizabilities of $3d_{\frac32m=\frac32}$ state 
have a big difference with dynamic polarizabilities for the linearly polarized light.
When the wavelength is close to $3d_{\frac32}-4p_{\frac32}$ and $3d_{\frac32}-4p_{\frac12}$ transition
wavelengths, $\alpha_{3d_{\frac32m=\frac32}} \to \infty$ for linearly polarized light.
The dynamic polarizabilities of $3d_{\frac32m=\frac32}$ state are not infinity 
for the right handed circularly polarized light.
The $3d_{\frac32}-4p_{\frac12}$ or $3d_{\frac32}-4p_{\frac32}$ transition have no 
contribution to the polarizability 
when the photo energy is close to their transition energy.
The scalar, vector and tensor polarizabilities offset each other, 
\begin{eqnarray}
\label{z}
& & \lim_{\omega \rightarrow \Delta E(3d_{\frac32}-np_j)} 
(\frac{1}{3(2j_{i}+1)}|\langle\psi_{3d_{\frac32}}\|D\|\psi_{np_j}\rangle|^{2} \nonumber \\
&\times&\left[ \frac{1}{\Delta E_{3d_{\frac32}-np_j}+\omega}+\frac{1}{\Delta E_{3d_{\frac32}-np_j}-\omega}\right] \nonumber \\
& - & A\frac{m_{j_i}}{2j_i} \sqrt {\frac{6j_{i}}{(j_i+1)(2j_i+1)}} 
\left \{
\begin{array}{ccc}
j_i & 1    & j_n \\
1   & j_i  & 1 \\
\end {array}
\right \}\nonumber \\
&\times& (-1)^{j_i +j_n +1}|\langle\psi_{4s}\|D\|\psi_{np_j}\rangle|^{2} \nonumber \\ 
&\times&\left[ \frac{1}{\Delta E_{3d_{\frac32}-np_j}+\omega}-\frac{1}{\Delta E_{3d_{\frac32}-np_j}-\omega}\right] \nonumber \\
&- & \frac{3m_{j_i}^{2}-j_i(j_i+1)}{2j_i(2j_i-1)} \times (-2)\sqrt{\frac{5j_i(2j_i-1)}{6(j_i+1)(2j_i+1)(2j_i+3)}}    \nonumber \\
& \times & \left \{
\begin{array}{ccc}
j_i & 1    & j_n \\
1   & j_i  & 2 \\
\end {array}
\right \} \times (-1)^{j_i +j_n +1}|\langle\psi_{3d_{\frac32}}\|D\|\psi_{np_j}\rangle|^{2} \nonumber \\
&\times&\left[ \frac{1}{\Delta E_{3d_{\frac32}-np_j}+\omega}+\frac{1}{\Delta E_{3d_{\frac32}-np_j}-\omega}\right] ) \nonumber \\
&=&  0. 
\end{eqnarray}
where, $j_i$ = $\frac32$ is total angular momentum for $3d_{\frac32}$ state,
$j_n$ = $\frac12,\frac32$ is total angular momentum for $np_j$ state
and $m_{j_i}$ = $+\frac32$ is magnetic quantum number for $3d_{\frac32}$ state.
The magic wavelengths are listed in Table \ref{tab4}.
It can be found that there are two or three magic 
wavelength for each of $4s_{\frac12m=-\frac12}-3d_{\frac32,m}$ transitions. 
It should be noted that wavelength near 394.6 nm lies
between the $4s-4p_j$ resonant transitions which is smaller about 1 nm than 394.79 nm
of $4s-3d_{\frac32m}$ transition for the linearly polarized light. 
The measurement of magic wavelength near 394.6 nm can be use to a 
further check the ratio of 
the $f_{4s-4p_j}$ oscillator strength. 
As mentioned before, the measurement of magic wavelength near 850 nm 
for $4s_{\frac12m=-\frac12}-3d_{\frac32,m=\pm\frac12}$
is able to determine the oscillator strength ratio of
$f_{4p_{\frac32} \to 3d_{\frac32}} : f_{4p_{\frac32} \to 3d_{\frac52}}$.  
 
The dynamic polarizabilities of the $4s_{\frac12m=-\frac12}$  and $3d_{\frac52,m}$ states are shown in Fig.\ref{fig5}.
Table \ref{tab5} gives the magic wavelengths of $4s_{\frac12m=-\frac12}  - 3d_{\frac52,m}$ transitions.  
There are only one or two magic 
wavelength for each of $4s_{\frac12m=-\frac12}-3d_{\frac32,m}$ transition. 
Similar to the experiment of Liu \emph{et al.}\cite{liu15a},
the measurement of magic wavelength near 394.6 nm can be use to a 
further check the ratio of 
the $f_{4s-4p_j}$ oscillator strength. 
                                                                                                                                                                                                                                                                                                                                                                                                                                                                               
\section{Conclusions}  

The energy levels, electric dipole matrix elements 
and static polarizabilities are calculated using 
RCICP procedure. The polarizabilties for the $4s$, 
$4p_j$ and $3d_j$ states agree with the available theoretical and 
experimental results very well. The BBR shift for the $4s-3d_{\frac52}$ 
clock transition is also determined and the present result is within the recent 
experimental error bar\cite{gao16a}.

The dynamic dipole polarizabilities of the
$4s$, $4p_{j}$ and $3d_{j}$ states of Ca$^{+}$ ion are calculated for
the linearly and circularly polarized light.
The magic wavelengths for each of magnetic sublevels of  $4s-4p_j$ and $4s-3d_j$ transitions are determined. 
The magic wavelengths for the linearly polarized light 
agree with the available theoretical results\cite{tang13b,kaur15a} excellently.
The magic wavelengths for the circularly polarized light have very big difference with 
those for linearly polarized light. Some additional magic wavelength are found for circularly 
polarized light. We recommend that the measurement of the magic wavelength near 850 nm of 
$4s-4p_{\frac32,m=\pm\frac32,\pm\frac12}$ could be able to determine the oscillator strength ratio of
$f_{4p_{\frac32} \to 3d_{\frac32}}$ and $f_{4p_{\frac32} \to 3d_{\frac52}}$.

\section{Acknowledgments}

The work of JJ was supported by National Natural Science Foundation
of China (NSFC) (Grants No.11564036).
The work of LYX was supported by NSFC(Grants No. U1331122).
The work of DHZ was supported by NSFC (Grants No. 11464042,U1330117). The
work of CZD was supported by NSFC (Grants No.11274254, U1332206).


\begin{thebibliography}{44}
\expandafter\ifx\csname natexlab\endcsname\relax\def\natexlab#1{#1}\fi
\expandafter\ifx\csname bibnamefont\endcsname\relax
  \def\bibnamefont#1{#1}\fi
\expandafter\ifx\csname bibfnamefont\endcsname\relax
  \def\bibfnamefont#1{#1}\fi
\expandafter\ifx\csname citenamefont\endcsname\relax
  \def\citenamefont#1{#1}\fi
\expandafter\ifx\csname url\endcsname\relax
  \def\url#1{\texttt{#1}}\fi
\expandafter\ifx\csname urlprefix\endcsname\relax\def\urlprefix{URL }\fi
\providecommand{\bibinfo}[2]{#2}
\providecommand{\eprint}[2][]{\url{#2}}

\bibitem[{\citenamefont{Ye et~al.}(1999)\citenamefont{Ye, Vernooy, and
  Kimble}}]{ye99a}
\bibinfo{author}{\bibfnamefont{J.}~\bibnamefont{Ye}},
  \bibinfo{author}{\bibfnamefont{D.~W.} \bibnamefont{Vernooy}},
  \bibnamefont{and} \bibinfo{author}{\bibfnamefont{H.~J.}
  \bibnamefont{Kimble}}, \bibinfo{journal}{Phys. Rev. Lett.}
  \textbf{\bibinfo{volume}{83}}, \bibinfo{pages}{4987} (\bibinfo{year}{1999}).

\bibitem[{\citenamefont{{Katori} et~al.}(1999)\citenamefont{{Katori}, {Ido},
  and {Kuwata-Gonokami}}}]{katori99b}
\bibinfo{author}{\bibfnamefont{H.}~\bibnamefont{{Katori}}},
  \bibinfo{author}{\bibfnamefont{T.}~\bibnamefont{{Ido}}}, \bibnamefont{and}
  \bibinfo{author}{\bibfnamefont{M.}~\bibnamefont{{Kuwata-Gonokami}}},
  \bibinfo{journal}{J.~Phys.~Soc.~Japan} \textbf{\bibinfo{volume}{68}},
  \bibinfo{pages}{2479} (\bibinfo{year}{1999}).

\bibitem[{\citenamefont{Takamoto and Katori}(2003)}]{takamoto03a}
\bibinfo{author}{\bibfnamefont{M.}~\bibnamefont{Takamoto}} \bibnamefont{and}
  \bibinfo{author}{\bibfnamefont{H.}~\bibnamefont{Katori}},
  \bibinfo{journal}{Phys. Rev. Lett.} \textbf{\bibinfo{volume}{91}},
  \bibinfo{pages}{223001} (\bibinfo{year}{2003}).

\bibitem[{\citenamefont{{Bauch}}(2003)}]{bauch03a}
\bibinfo{author}{\bibfnamefont{A.}~\bibnamefont{{Bauch}}},
  \bibinfo{journal}{Meas.~Sci.~Technol.} \textbf{\bibinfo{volume}{14}},
  \bibinfo{pages}{1159} (\bibinfo{year}{2003}).

\bibitem[{\citenamefont{{Gill} et~al.}(2003)\citenamefont{{Gill}, {Barwood},
  {Klein}, {Huang}, {Webster}, {Blythe}, {Hosaka}, {Lea}, and
  {Margolis}}}]{gill03a}
\bibinfo{author}{\bibfnamefont{P.}~\bibnamefont{{Gill}}},
  \bibinfo{author}{\bibfnamefont{G.~P.} \bibnamefont{{Barwood}}},
  \bibinfo{author}{\bibfnamefont{H.~A.} \bibnamefont{{Klein}}},
  \bibinfo{author}{\bibfnamefont{G.}~\bibnamefont{{Huang}}},
  \bibinfo{author}{\bibfnamefont{S.~A.} \bibnamefont{{Webster}}},
  \bibinfo{author}{\bibfnamefont{P.~J.} \bibnamefont{{Blythe}}},
  \bibinfo{author}{\bibfnamefont{K.}~\bibnamefont{{Hosaka}}},
  \bibinfo{author}{\bibfnamefont{S.~N.} \bibnamefont{{Lea}}}, \bibnamefont{and}
  \bibinfo{author}{\bibfnamefont{H.~S.} \bibnamefont{{Margolis}}},
  \bibinfo{journal}{Meas.~Sci.~Technol.} \textbf{\bibinfo{volume}{14}},
  \bibinfo{pages}{1174} (\bibinfo{year}{2003}).

\bibitem[{\citenamefont{{Gill}}(2005)}]{gill05a}
\bibinfo{author}{\bibfnamefont{P.}~\bibnamefont{{Gill}}},
  \bibinfo{journal}{Metrologia} \textbf{\bibinfo{volume}{42}},
  \bibinfo{pages}{S125} (\bibinfo{year}{2005}).

\bibitem[{\citenamefont{{Lorini} et~al.}(2008)\citenamefont{{Lorini}, {Ashby},
  {Brusch}, {Diddams}, {Drullinger}, {Eason}, {Fortier}, {Hastings}, {Heavner},
  {Hume} et~al.}}]{lorini08a}
\bibinfo{author}{\bibfnamefont{L.}~\bibnamefont{{Lorini}}},
  \bibinfo{author}{\bibfnamefont{N.}~\bibnamefont{{Ashby}}},
  \bibinfo{author}{\bibfnamefont{A.}~\bibnamefont{{Brusch}}},
  \bibinfo{author}{\bibfnamefont{S.}~\bibnamefont{{Diddams}}},
  \bibinfo{author}{\bibfnamefont{R.}~\bibnamefont{{Drullinger}}},
  \bibinfo{author}{\bibfnamefont{E.}~\bibnamefont{{Eason}}},
  \bibinfo{author}{\bibfnamefont{T.}~\bibnamefont{{Fortier}}},
  \bibinfo{author}{\bibfnamefont{P.}~\bibnamefont{{Hastings}}},
  \bibinfo{author}{\bibfnamefont{T.}~\bibnamefont{{Heavner}}},
  \bibinfo{author}{\bibfnamefont{D.}~\bibnamefont{{Hume}}},
  \bibnamefont{et~al.}, \bibinfo{journal}{Eur.~Phys.~J.~Special Topics}
  \textbf{\bibinfo{volume}{163}}, \bibinfo{pages}{19} (\bibinfo{year}{2008}).

\bibitem[{\citenamefont{{Gill}}(2011)}]{gill11a}
\bibinfo{author}{\bibfnamefont{P.}~\bibnamefont{{Gill}}},
  \bibinfo{journal}{Royal Soc. of London Phil.~Trans. Series A}
  \textbf{\bibinfo{volume}{369}}, \bibinfo{pages}{4109} (\bibinfo{year}{2011}).

\bibitem[{\citenamefont{Kirchmair et~al.}(2009)\citenamefont{Kirchmair,
  Benhelm, Z\"ahringer, Gerritsma, Roos, and Blatt}}]{kirchmair09a}
\bibinfo{author}{\bibfnamefont{G.}~\bibnamefont{Kirchmair}},
  \bibinfo{author}{\bibfnamefont{J.}~\bibnamefont{Benhelm}},
  \bibinfo{author}{\bibfnamefont{F.}~\bibnamefont{Z\"ahringer}},
  \bibinfo{author}{\bibfnamefont{R.}~\bibnamefont{Gerritsma}},
  \bibinfo{author}{\bibfnamefont{C.~F.} \bibnamefont{Roos}}, \bibnamefont{and}
  \bibinfo{author}{\bibfnamefont{R.}~\bibnamefont{Blatt}},
  \bibinfo{journal}{Phys. Rev. A} \textbf{\bibinfo{volume}{79}},
  \bibinfo{pages}{020304} (\bibinfo{year}{2009}).

\bibitem[{\citenamefont{Sahoo and Arora}(2013)}]{sahoo13a}
\bibinfo{author}{\bibfnamefont{B.~K.} \bibnamefont{Sahoo}} \bibnamefont{and}
  \bibinfo{author}{\bibfnamefont{B.}~\bibnamefont{Arora}},
  \bibinfo{journal}{Phys. Rev. A} \textbf{\bibinfo{volume}{87}},
  \bibinfo{pages}{023402} (\bibinfo{year}{2013}).

\bibitem[{\citenamefont{Wilpers et~al.}(2007)\citenamefont{Wilpers, Oates,
  Diddams, Bartels, Fortier, Oskay, Bergquist, Jefferts, Heavner, Parker
  et~al.}}]{wilpers07a}
\bibinfo{author}{\bibfnamefont{G.}~\bibnamefont{Wilpers}},
  \bibinfo{author}{\bibfnamefont{C.~W.} \bibnamefont{Oates}},
  \bibinfo{author}{\bibfnamefont{S.~A.} \bibnamefont{Diddams}},
  \bibinfo{author}{\bibfnamefont{A.}~\bibnamefont{Bartels}},
  \bibinfo{author}{\bibfnamefont{T.~M.} \bibnamefont{Fortier}},
  \bibinfo{author}{\bibfnamefont{W.~H.} \bibnamefont{Oskay}},
  \bibinfo{author}{\bibfnamefont{J.~C.} \bibnamefont{Bergquist}},
  \bibinfo{author}{\bibfnamefont{S.~R.} \bibnamefont{Jefferts}},
  \bibinfo{author}{\bibfnamefont{T.~P.} \bibnamefont{Heavner}},
  \bibinfo{author}{\bibfnamefont{T.~E.} \bibnamefont{Parker}},
  \bibnamefont{et~al.}, \bibinfo{journal}{Metrologia}
  \textbf{\bibinfo{volume}{44}}, \bibinfo{pages}{146} (\bibinfo{year}{2007}).

\bibitem[{\citenamefont{Ludlow et~al.}(2008)\citenamefont{Ludlow, Zelevinsky,
  Campbell, Blatt, Boyd, de~Miranda, Martin, Thomsen, Foreman, and
  Ye}}]{ludlow08a}
\bibinfo{author}{\bibfnamefont{A.~D.} \bibnamefont{Ludlow}},
  \bibinfo{author}{\bibfnamefont{T.}~\bibnamefont{Zelevinsky}},
  \bibinfo{author}{\bibfnamefont{G.~K.} \bibnamefont{Campbell}},
  \bibinfo{author}{\bibfnamefont{S.}~\bibnamefont{Blatt}},
  \bibinfo{author}{\bibfnamefont{M.~M.} \bibnamefont{Boyd}},
  \bibinfo{author}{\bibfnamefont{M.~H.} \bibnamefont{de~Miranda}},
  \bibinfo{author}{\bibfnamefont{M.~J.} \bibnamefont{Martin}},
  \bibinfo{author}{\bibfnamefont{J.~W.} \bibnamefont{Thomsen}},
  \bibinfo{author}{\bibfnamefont{S.~M.} \bibnamefont{Foreman}},
  \bibnamefont{and} \bibinfo{author}{\bibfnamefont{J.}~\bibnamefont{Ye}},
  \bibinfo{journal}{Science} \textbf{\bibinfo{volume}{319}},
  \bibinfo{pages}{1805} (\bibinfo{year}{2008}).

\bibitem[{\citenamefont{Kaur et~al.}(2015)\citenamefont{Kaur, Singh, Arora, and
  Sahoo}}]{kaur15a}
\bibinfo{author}{\bibfnamefont{J.}~\bibnamefont{Kaur}},
  \bibinfo{author}{\bibfnamefont{S.}~\bibnamefont{Singh}},
  \bibinfo{author}{\bibfnamefont{B.}~\bibnamefont{Arora}}, \bibnamefont{and}
  \bibinfo{author}{\bibfnamefont{B.~K.} \bibnamefont{Sahoo}},
  \bibinfo{journal}{Phys. Rev. A} \textbf{\bibinfo{volume}{92}},
  \bibinfo{pages}{031402} (\bibinfo{year}{2015}).

\bibitem[{\citenamefont{Liu et~al.}(2015)\citenamefont{Liu, Huang, Bian, Shao,
  Guan, Tang, Li, Mitroy, and Gao}}]{liu15a}
\bibinfo{author}{\bibfnamefont{P.-L.} \bibnamefont{Liu}},
  \bibinfo{author}{\bibfnamefont{Y.}~\bibnamefont{Huang}},
  \bibinfo{author}{\bibfnamefont{W.}~\bibnamefont{Bian}},
  \bibinfo{author}{\bibfnamefont{H.}~\bibnamefont{Shao}},
  \bibinfo{author}{\bibfnamefont{H.}~\bibnamefont{Guan}},
  \bibinfo{author}{\bibfnamefont{Y.-B.} \bibnamefont{Tang}},
  \bibinfo{author}{\bibfnamefont{C.-B.} \bibnamefont{Li}},
  \bibinfo{author}{\bibfnamefont{J.}~\bibnamefont{Mitroy}}, \bibnamefont{and}
  \bibinfo{author}{\bibfnamefont{K.-L.} \bibnamefont{Gao}},
  \bibinfo{journal}{Phys. Rev. Lett.} \textbf{\bibinfo{volume}{114}},
  \bibinfo{pages}{223001} (\bibinfo{year}{2015}).

\bibitem[{\citenamefont{Tang et~al.}(2013)\citenamefont{Tang, Qiao, Shi, and
  Mitroy}}]{tang13b}
\bibinfo{author}{\bibfnamefont{Y.-B.} \bibnamefont{Tang}},
  \bibinfo{author}{\bibfnamefont{H.-X.} \bibnamefont{Qiao}},
  \bibinfo{author}{\bibfnamefont{T.-Y.} \bibnamefont{Shi}}, \bibnamefont{and}
  \bibinfo{author}{\bibfnamefont{J.}~\bibnamefont{Mitroy}},
  \bibinfo{journal}{Phys. Rev. A} \textbf{\bibinfo{volume}{87}},
  \bibinfo{pages}{042517} (\bibinfo{year}{2013}).

\bibitem[{\citenamefont{{Lundblad} et~al.}(2010)\citenamefont{{Lundblad},
  {Schlosser}, and {Porto}}}]{lundblad10a}
\bibinfo{author}{\bibfnamefont{N.}~\bibnamefont{{Lundblad}}},
  \bibinfo{author}{\bibfnamefont{M.}~\bibnamefont{{Schlosser}}},
  \bibnamefont{and} \bibinfo{author}{\bibfnamefont{J.~V.}
  \bibnamefont{{Porto}}}, \bibinfo{journal}{\pra}
  \textbf{\bibinfo{volume}{81}}, \bibinfo{eid}{031611} (\bibinfo{year}{2010}).

\bibitem[{\citenamefont{Arora et~al.}(2007)\citenamefont{Arora, Safronova, and
  Clark}}]{arora07a}
\bibinfo{author}{\bibfnamefont{B.}~\bibnamefont{Arora}},
  \bibinfo{author}{\bibfnamefont{M.~S.} \bibnamefont{Safronova}},
  \bibnamefont{and} \bibinfo{author}{\bibfnamefont{C.~W.} \bibnamefont{Clark}},
  \bibinfo{journal}{Phys. Rev. A} \textbf{\bibinfo{volume}{76}},
  \bibinfo{pages}{064501} (\bibinfo{year}{2007}).

\bibitem[{\citenamefont{Singh et~al.}(2016{\natexlab{a}})\citenamefont{Singh,
  Sahoo, and Arora}}]{singh16b}
\bibinfo{author}{\bibfnamefont{S.}~\bibnamefont{Singh}},
  \bibinfo{author}{\bibfnamefont{B.~K.} \bibnamefont{Sahoo}}, \bibnamefont{and}
  \bibinfo{author}{\bibfnamefont{B.}~\bibnamefont{Arora}},
  \bibinfo{journal}{Phys. Rev. A} \textbf{\bibinfo{volume}{94}},
  \bibinfo{pages}{023418} (\bibinfo{year}{2016}{\natexlab{a}}).

\bibitem[{\citenamefont{Arora and Sahoo}(2012)}]{arora12a}
\bibinfo{author}{\bibfnamefont{B.}~\bibnamefont{Arora}} \bibnamefont{and}
  \bibinfo{author}{\bibfnamefont{B.~K.} \bibnamefont{Sahoo}},
  \bibinfo{journal}{Phys. Rev. A} \textbf{\bibinfo{volume}{86}},
  \bibinfo{pages}{033416} (\bibinfo{year}{2012}).

\bibitem[{\citenamefont{Singh et~al.}(2016{\natexlab{b}})\citenamefont{Singh,
  Sahoo, and Arora}}]{singh16a}
\bibinfo{author}{\bibfnamefont{S.}~\bibnamefont{Singh}},
  \bibinfo{author}{\bibfnamefont{B.~K.} \bibnamefont{Sahoo}}, \bibnamefont{and}
  \bibinfo{author}{\bibfnamefont{B.}~\bibnamefont{Arora}},
  \bibinfo{journal}{Phys. Rev. A} \textbf{\bibinfo{volume}{93}},
  \bibinfo{pages}{063422} (\bibinfo{year}{2016}{\natexlab{b}}).

\bibitem[{\citenamefont{Singh et~al.}(2016{\natexlab{c}})\citenamefont{Singh,
  Kaur, Sahoo, and Arora}}]{cs16a}
\bibinfo{author}{\bibfnamefont{S.}~\bibnamefont{Singh}},
  \bibinfo{author}{\bibfnamefont{K.}~\bibnamefont{Kaur}},
  \bibinfo{author}{\bibfnamefont{B.~K.} \bibnamefont{Sahoo}}, \bibnamefont{and}
  \bibinfo{author}{\bibfnamefont{B.}~\bibnamefont{Arora}},
  \bibinfo{journal}{Journal of Physics B Atomic Molecular Physics}
  \textbf{\bibinfo{volume}{49}} (\bibinfo{year}{2016}{\natexlab{c}}).

\bibitem[{\citenamefont{Le~Kien et~al.}(2013)\citenamefont{Le~Kien,
  Schneeweiss, and Rauschenbeutel}}]{kien13a}
\bibinfo{author}{\bibfnamefont{F.}~\bibnamefont{Le~Kien}},
  \bibinfo{author}{\bibfnamefont{P.}~\bibnamefont{Schneeweiss}},
  \bibnamefont{and}
  \bibinfo{author}{\bibfnamefont{A.}~\bibnamefont{Rauschenbeutel}},
  \bibinfo{journal}{The European Physical Journal D}
  \textbf{\bibinfo{volume}{67}}, \bibinfo{eid}{92} (\bibinfo{year}{2013}), ISSN
  \bibinfo{issn}{1434-6060}.

\bibitem[{\citenamefont{Gao}(2013)}]{kelin13a}
\bibinfo{author}{\bibfnamefont{K.~L.} \bibnamefont{Gao}},
  \bibinfo{journal}{Science Bulletin} \textbf{\bibinfo{volume}{58}},
  \bibinfo{pages}{853} (\bibinfo{year}{2013}).

\bibitem[{\citenamefont{Huang et~al.}(2012)\citenamefont{Huang, Cao, Liu,
  Liang, Ou, Guan, Huang, Li, and Gao}}]{hertz12a}
\bibinfo{author}{\bibfnamefont{Y.}~\bibnamefont{Huang}},
  \bibinfo{author}{\bibfnamefont{J.}~\bibnamefont{Cao}},
  \bibinfo{author}{\bibfnamefont{P.}~\bibnamefont{Liu}},
  \bibinfo{author}{\bibfnamefont{K.}~\bibnamefont{Liang}},
  \bibinfo{author}{\bibfnamefont{B.}~\bibnamefont{Ou}},
  \bibinfo{author}{\bibfnamefont{H.}~\bibnamefont{Guan}},
  \bibinfo{author}{\bibfnamefont{X.}~\bibnamefont{Huang}},
  \bibinfo{author}{\bibfnamefont{T.}~\bibnamefont{Li}}, \bibnamefont{and}
  \bibinfo{author}{\bibfnamefont{K.}~\bibnamefont{Gao}},
  \bibinfo{journal}{Phys. Rev. A} \textbf{\bibinfo{volume}{85}},
  \bibinfo{pages}{030503} (\bibinfo{year}{2012}).

\bibitem[{\citenamefont{Degenhardt et~al.}(2004)\citenamefont{Degenhardt,
  Stoehr, Sterr, Riehle, and Lisdat}}]{degenhardt04a}
\bibinfo{author}{\bibfnamefont{C.}~\bibnamefont{Degenhardt}},
  \bibinfo{author}{\bibfnamefont{H.}~\bibnamefont{Stoehr}},
  \bibinfo{author}{\bibfnamefont{U.}~\bibnamefont{Sterr}},
  \bibinfo{author}{\bibfnamefont{F.}~\bibnamefont{Riehle}}, \bibnamefont{and}
  \bibinfo{author}{\bibfnamefont{C.}~\bibnamefont{Lisdat}},
  \bibinfo{journal}{Phys. Rev. A} \textbf{\bibinfo{volume}{70}},
  \bibinfo{pages}{023414} (\bibinfo{year}{2004}).

\bibitem[{\citenamefont{Chwalla et~al.}(2009)\citenamefont{Chwalla, Benhelm,
  Kim, Kirchmair, Monz, Riebe, Schindler, Villar, H\"ansel, Roos
  et~al.}}]{chwalla09a}
\bibinfo{author}{\bibfnamefont{M.}~\bibnamefont{Chwalla}},
  \bibinfo{author}{\bibfnamefont{J.}~\bibnamefont{Benhelm}},
  \bibinfo{author}{\bibfnamefont{K.}~\bibnamefont{Kim}},
  \bibinfo{author}{\bibfnamefont{G.}~\bibnamefont{Kirchmair}},
  \bibinfo{author}{\bibfnamefont{T.}~\bibnamefont{Monz}},
  \bibinfo{author}{\bibfnamefont{M.}~\bibnamefont{Riebe}},
  \bibinfo{author}{\bibfnamefont{P.}~\bibnamefont{Schindler}},
  \bibinfo{author}{\bibfnamefont{A.~S.} \bibnamefont{Villar}},
  \bibinfo{author}{\bibfnamefont{W.}~\bibnamefont{H\"ansel}},
  \bibinfo{author}{\bibfnamefont{C.~F.} \bibnamefont{Roos}},
  \bibnamefont{et~al.}, \bibinfo{journal}{Phys. Rev. Lett.}
  \textbf{\bibinfo{volume}{102}}, \bibinfo{pages}{023002}
  (\bibinfo{year}{2009}).

\bibitem[{\citenamefont{Zhang et~al.}(2017)\citenamefont{Zhang, Cao, lin Shu,
  bo~Yuan, juan Shang, feng Cui, jia Chao, mao Wang, xin Liu, and ren
  Huang}}]{zhang16a}
\bibinfo{author}{\bibfnamefont{P.}~\bibnamefont{Zhang}},
  \bibinfo{author}{\bibfnamefont{J.}~\bibnamefont{Cao}},
  \bibinfo{author}{\bibfnamefont{H.}~\bibnamefont{lin Shu}},
  \bibinfo{author}{\bibfnamefont{J.}~\bibnamefont{bo~Yuan}},
  \bibinfo{author}{\bibfnamefont{J.}~\bibnamefont{juan Shang}},
  \bibinfo{author}{\bibfnamefont{K.}~\bibnamefont{feng Cui}},
  \bibinfo{author}{\bibfnamefont{S.}~\bibnamefont{jia Chao}},
  \bibinfo{author}{\bibfnamefont{S.}~\bibnamefont{mao Wang}},
  \bibinfo{author}{\bibfnamefont{D.}~\bibnamefont{xin Liu}}, \bibnamefont{and}
  \bibinfo{author}{\bibfnamefont{X.}~\bibnamefont{ren Huang}},
  \bibinfo{journal}{Journal of Physics B: Atomic, Molecular and Optical
  Physics} \textbf{\bibinfo{volume}{50}}, \bibinfo{pages}{015002}
  (\bibinfo{year}{2017}).

\bibitem[{\citenamefont{Häffner et~al.}(2008)\citenamefont{Häffner, Roos, and
  Blatt}}]{haff08a}
\bibinfo{author}{\bibfnamefont{H.}~\bibnamefont{Häffner}},
  \bibinfo{author}{\bibfnamefont{C.}~\bibnamefont{Roos}}, \bibnamefont{and}
  \bibinfo{author}{\bibfnamefont{R.}~\bibnamefont{Blatt}},
  \bibinfo{journal}{Physics Reports} \textbf{\bibinfo{volume}{469}},
  \bibinfo{pages}{155 } (\bibinfo{year}{2008}).

\bibitem[{\citenamefont{Huang et~al.}(2016)\citenamefont{Huang, Guan, Liu,
  Bian, Ma, Liang, Li, and Gao}}]{huang16a}
\bibinfo{author}{\bibfnamefont{Y.}~\bibnamefont{Huang}},
  \bibinfo{author}{\bibfnamefont{H.}~\bibnamefont{Guan}},
  \bibinfo{author}{\bibfnamefont{P.}~\bibnamefont{Liu}},
  \bibinfo{author}{\bibfnamefont{W.}~\bibnamefont{Bian}},
  \bibinfo{author}{\bibfnamefont{L.}~\bibnamefont{Ma}},
  \bibinfo{author}{\bibfnamefont{K.}~\bibnamefont{Liang}},
  \bibinfo{author}{\bibfnamefont{T.}~\bibnamefont{Li}}, \bibnamefont{and}
  \bibinfo{author}{\bibfnamefont{K.}~\bibnamefont{Gao}},
  \bibinfo{journal}{Phys. Rev. Lett.} \textbf{\bibinfo{volume}{116}},
  \bibinfo{pages}{013001} (\bibinfo{year}{2016}).

\bibitem[{\citenamefont{Safronova and
  Safronova}(2011{\natexlab{a}})}]{safronova11a}
\bibinfo{author}{\bibfnamefont{M.~S.} \bibnamefont{Safronova}}
  \bibnamefont{and} \bibinfo{author}{\bibfnamefont{U.~I.}
  \bibnamefont{Safronova}}, \bibinfo{journal}{Phys. Rev. A}
  \textbf{\bibinfo{volume}{83}}, \bibinfo{pages}{012503}
  (\bibinfo{year}{2011}{\natexlab{a}}).

\bibitem[{\citenamefont{Mitroy et~al.}(2008)\citenamefont{Mitroy, Zhang,
  Bromley, and Young}}]{mitroy08a}
\bibinfo{author}{\bibfnamefont{J.}~\bibnamefont{Mitroy}},
  \bibinfo{author}{\bibfnamefont{J.~Y.} \bibnamefont{Zhang}},
  \bibinfo{author}{\bibfnamefont{M.~W.~J.} \bibnamefont{Bromley}},
  \bibnamefont{and} \bibinfo{author}{\bibfnamefont{S.~I.} \bibnamefont{Young}},
  \bibinfo{journal}{Phys. Rev. A} \textbf{\bibinfo{volume}{78}},
  \bibinfo{pages}{012715} (\bibinfo{year}{2008}).

\bibitem[{\citenamefont{Sahoo et~al.}(2009)\citenamefont{Sahoo, Das, and
  Mukherjee}}]{sahoo09a}
\bibinfo{author}{\bibfnamefont{B.~K.} \bibnamefont{Sahoo}},
  \bibinfo{author}{\bibfnamefont{B.~P.} \bibnamefont{Das}}, \bibnamefont{and}
  \bibinfo{author}{\bibfnamefont{D.}~\bibnamefont{Mukherjee}},
  \bibinfo{journal}{Phys. Rev. A} \textbf{\bibinfo{volume}{79}},
  \bibinfo{pages}{052511} (\bibinfo{year}{2009}).

\bibitem[{\citenamefont{Chang}(1983)}]{chang83a}
\bibinfo{author}{\bibfnamefont{E.~S.} \bibnamefont{Chang}},
  \bibinfo{journal}{Journal of Physics B: Atomic and Molecular Physics}
  \textbf{\bibinfo{volume}{16}}, \bibinfo{pages}{L539} (\bibinfo{year}{1983}).

\bibitem[{\citenamefont{{Champenois} et~al.}(2005)\citenamefont{{Champenois},
  {Knoop}, {Houssin}, {Hagel}, {Vedel}, and {Vedel}}}]{cham04a}
\bibinfo{author}{\bibfnamefont{C.}~\bibnamefont{{Champenois}}},
  \bibinfo{author}{\bibfnamefont{M.}~\bibnamefont{{Knoop}}},
  \bibinfo{author}{\bibfnamefont{M.}~\bibnamefont{{Houssin}}},
  \bibinfo{author}{\bibfnamefont{G.}~\bibnamefont{{Hagel}}},
  \bibinfo{author}{\bibfnamefont{M.}~\bibnamefont{{Vedel}}}, \bibnamefont{and}
  \bibinfo{author}{\bibfnamefont{F.}~\bibnamefont{{Vedel}}},
  \bibinfo{journal}{ArXiv Physics e-prints}  (\bibinfo{year}{2005}),
  \eprint{physics/0511089}.

\bibitem[{\citenamefont{Kajita et~al.}(2005)\citenamefont{Kajita, Li,
  Matsubara, Hayasaka, and Hosokawa}}]{kajita05a}
\bibinfo{author}{\bibfnamefont{M.}~\bibnamefont{Kajita}},
  \bibinfo{author}{\bibfnamefont{Y.}~\bibnamefont{Li}},
  \bibinfo{author}{\bibfnamefont{K.}~\bibnamefont{Matsubara}},
  \bibinfo{author}{\bibfnamefont{K.}~\bibnamefont{Hayasaka}}, \bibnamefont{and}
  \bibinfo{author}{\bibfnamefont{M.}~\bibnamefont{Hosokawa}},
  \bibinfo{journal}{Phys. Rev. A} \textbf{\bibinfo{volume}{72}},
  \bibinfo{pages}{043404} (\bibinfo{year}{2005}).

\bibitem[{\citenamefont{Guan et~al.}(2016)\citenamefont{Guan, Huang, and
  Gao}}]{gao16a}
\bibinfo{author}{\bibfnamefont{H.}~\bibnamefont{Guan}},
  \bibinfo{author}{\bibfnamefont{Y.}~\bibnamefont{Huang}}, \bibnamefont{and}
  \bibinfo{author}{\bibfnamefont{K.-L.} \bibnamefont{Gao}},
  \bibinfo{journal}{Scientia Sinica, Physica,Mechanica Astronomica}
  \textbf{\bibinfo{volume}{7}}, \bibinfo{pages}{006} (\bibinfo{year}{2016}).

\bibitem[{\citenamefont{{Huang} et~al.}(2012)\citenamefont{{Huang}, {Cao},
  {Liu}, {Liang}, {Ou}, {Guan}, {Huang}, {Li}, and {Gao}}}]{huang12a}
\bibinfo{author}{\bibfnamefont{Y.}~\bibnamefont{{Huang}}},
  \bibinfo{author}{\bibfnamefont{J.}~\bibnamefont{{Cao}}},
  \bibinfo{author}{\bibfnamefont{P.}~\bibnamefont{{Liu}}},
  \bibinfo{author}{\bibfnamefont{K.}~\bibnamefont{{Liang}}},
  \bibinfo{author}{\bibfnamefont{B.}~\bibnamefont{{Ou}}},
  \bibinfo{author}{\bibfnamefont{H.}~\bibnamefont{{Guan}}},
  \bibinfo{author}{\bibfnamefont{X.}~\bibnamefont{{Huang}}},
  \bibinfo{author}{\bibfnamefont{T.}~\bibnamefont{{Li}}}, \bibnamefont{and}
  \bibinfo{author}{\bibfnamefont{K.}~\bibnamefont{{Gao}}},
  \bibinfo{journal}{\pra} \textbf{\bibinfo{volume}{85}}, \bibinfo{eid}{030503}
  (\bibinfo{year}{2012}).

\bibitem[{\citenamefont{Huang et~al.}(2011)\citenamefont{Huang, Liu, Cao, Ou,
  Liu, Guan, Huang, and Gao}}]{yaohuang11a}
\bibinfo{author}{\bibfnamefont{Y.}~\bibnamefont{Huang}},
  \bibinfo{author}{\bibfnamefont{Q.}~\bibnamefont{Liu}},
  \bibinfo{author}{\bibfnamefont{J.}~\bibnamefont{Cao}},
  \bibinfo{author}{\bibfnamefont{B.}~\bibnamefont{Ou}},
  \bibinfo{author}{\bibfnamefont{P.}~\bibnamefont{Liu}},
  \bibinfo{author}{\bibfnamefont{H.}~\bibnamefont{Guan}},
  \bibinfo{author}{\bibfnamefont{X.}~\bibnamefont{Huang}}, \bibnamefont{and}
  \bibinfo{author}{\bibfnamefont{K.}~\bibnamefont{Gao}},
  \bibinfo{journal}{Phys. Rev. A} \textbf{\bibinfo{volume}{84}},
  \bibinfo{pages}{053841} (\bibinfo{year}{2011}).

\bibitem[{\citenamefont{Jiang et~al.}(2016)\citenamefont{Jiang, Mitroy, Cheng,
  and Bromley}}]{jiang16a}
\bibinfo{author}{\bibfnamefont{J.}~\bibnamefont{Jiang}},
  \bibinfo{author}{\bibfnamefont{J.}~\bibnamefont{Mitroy}},
  \bibinfo{author}{\bibfnamefont{Y.}~\bibnamefont{Cheng}}, \bibnamefont{and}
  \bibinfo{author}{\bibfnamefont{M.~W.~J.} \bibnamefont{Bromley}},
  \bibinfo{journal}{Phys. Rev. A} \textbf{\bibinfo{volume}{94}},
  \bibinfo{pages}{062514} (\bibinfo{year}{2016}).

\bibitem[{\citenamefont{Safronova and
  Safronova}(2011{\natexlab{b}})}]{safronvoa11a}
\bibinfo{author}{\bibfnamefont{M.~S.} \bibnamefont{Safronova}}
  \bibnamefont{and} \bibinfo{author}{\bibfnamefont{U.~I.}
  \bibnamefont{Safronova}}, \bibinfo{journal}{Phys. Rev. A}
  \textbf{\bibinfo{volume}{83}}, \bibinfo{pages}{012503}
  (\bibinfo{year}{2011}{\natexlab{b}}).

\bibitem[{sup()}]{suppmat}
\bibinfo{journal}{See Supplemental Material at http://XXXXXXXX for additional
  tables of energy levels and matrix elements and breakdowns of
  polarizabilities at the magic wavelengths.}  (????).

\bibitem[{\citenamefont{Manakov et~al.}(1986)\citenamefont{Manakov,
  Ovsiannikov, and Rapoport}}]{nl86a}
\bibinfo{author}{\bibfnamefont{N.~L.} \bibnamefont{Manakov}},
  \bibinfo{author}{\bibfnamefont{V.~D.} \bibnamefont{Ovsiannikov}},
  \bibnamefont{and} \bibinfo{author}{\bibfnamefont{L.~P.}
  \bibnamefont{Rapoport}}, \bibinfo{journal}{Physics Reports}
  \textbf{\bibinfo{volume}{141}}, \bibinfo{pages}{320} (\bibinfo{year}{1986}).

\bibitem[{\citenamefont{Beloy}(2009)}]{beloy09a}
\bibinfo{author}{\bibfnamefont{K.}~\bibnamefont{Beloy}},
  \bibinfo{journal}{Ph.~D thesis,University of Nevada, Reno}
  (\bibinfo{year}{2009}).

\bibitem[{\citenamefont{Porsev and Derevianko}(2006)}]{porsev06a}
\bibinfo{author}{\bibfnamefont{S.~G.} \bibnamefont{Porsev}} \bibnamefont{and}
  \bibinfo{author}{\bibfnamefont{A.}~\bibnamefont{Derevianko}},
  \bibinfo{journal}{Phys. Rev. A} \textbf{\bibinfo{volume}{74}},
  \bibinfo{pages}{020502} (\bibinfo{year}{2006}).

\end{thebibliography}
\end{document}